\documentclass[12pt]{article}

\usepackage{amsmath}
\usepackage{amssymb}

\usepackage{indentfirst}
\usepackage{amssymb}
\usepackage{amsfonts}
\usepackage{amscd}
\usepackage{amsbsy}
\usepackage{amsthm}
\usepackage{latexsym}
\usepackage{graphicx,color} 
\usepackage[dvipsnames]{xcolor}
\usepackage[colorlinks]{hyperref}
\hypersetup{linkcolor=blue,citecolor=blue,urlcolor=blue}
\def\nn{\nonumber}       
\def\beq{\begin{eqnarray}}
\def\eeq{\end{eqnarray}}
\def\ln{\,\mbox{ln}\,}

\DeclareMathOperator{\cx}{\square}

\def\al{\alpha}
\def\be{\beta}

\def\ga{\gamma}
\def\de{\delta}
\def\vp{\varepsilon}
\def\ep{\epsilon}
\def\ze{\zeta}

\def\ka{\kappa}
\def\la{\lambda}

\def\pa{\partial}

\def\rh{\rho}
\def\si{\sigma}
\def\om{\omega}

\def\th{\theta}

\def\Ga{\Gamma}
\def\De{\Delta}
\def\La{\Lambda}
\def\Si{\Sigma}

\usepackage[symbol]{footmisc} 
\usepackage{float} 
\usepackage{cite}  

\begin{document}
	
\begin{center}
\renewcommand*{\thefootnote}{\fnsymbol{footnote}} 
{\Large \bf
Effective approach to the Antoniadis-Mottola model:
\\
quantum decoupling of the higher derivative terms}
\vskip 6mm

{\large \bf Wagno Cesar e Silva}
\footnote{E-mail address: \ wagnorion@gmail.com}
\quad
and
\quad
\ {\large \bf Ilya L. Shapiro}
\hspace{-1mm}\footnote{E-mail address: \ ilyashapiro2003@ufjf.br}
\vskip 6mm

Departamento de F\'{\i}sica, ICE, Universidade Federal de Juiz de Fora,
\\
36036-900, Juiz de Fora, Minas Gerais, Brazil
\end{center}
\vskip 2mm
\vskip 2mm


\begin{abstract}
	
\noindent
We explore the decoupling of massive ghost mode in the $4D$
(four-dimensional) theory of the conformal factor of the metric. The
model was introduced by Antoniadis and Mottola in \cite{antmot}
and can be regarded as a close analog of the fourth-derivative
quantum gravity. The analysis of the derived one-loop nonlocal form
factors includes their asymptotic behavior in the UV and IR limits.
In the UV (high energy) domain, our results reproduce the Minimal
Subtraction scheme-based beta functions of \cite{antmot}. In the IR
(i.e., at low energies), the diagrams with massive ghost internal
lines collapse into tadpole-type graphs without nonlocal
contributions and become irrelevant. On the other hand, those
structures that contribute to the running of parameters of the action
and survive in the IR, are well-correlated with the divergent part
(or the leading in UV contributions to the form factors), coming
from the effective low-energy theory of the conformal factor. This
effective theory describes only the light propagating mode. Finally,
we discuss whether these results may shed light on the possible
running of the cosmological constant at low energies.
\vskip 3mm


\end{abstract}

\setcounter{footnote}{0} 
\renewcommand*{\thefootnote}{\arabic{footnote}} 
\section{Introduction}
\label{sec0}

The running of the cosmological constant at low energies represents
an interesting alternative to the numerous models of Dark Energy,
as it provides the equation of state which is close, but not identical
to the $\om_\La = -1$, of the cosmological constant. On the other hand,
there is no full understanding of whether such a running is possible
or not, such that this issue remains uncertain and is a subject of
phenomenological considerations, as discussed in \cite{apco,DCCrun}
and many subsequent works. The main difficulty
for the thorough theoretical investigation is that the traditional
approach to quantum decoupling \cite{AC} implies calculating the
nonlocal form factor (or its equivalent) and taking its low-energy
limit. The cosmological constant acquires physical sense only in
curved spacetime and, in principle, the corresponding form factors
have to be built from covariant elements and analysed in curved
space. According to the Appelquist-Carazzone theorem \cite{AC},
heavy degrees of freedom decouple in the IR regime, and their loop
corrections are quadratically suppressed. The same effect should hold
in curved spacetime, leading to the corresponding decoupling
theorems.

The described program has been fulfilled in a series of papers
\cite{apco,fervi,Codello,Omar-FF4D} where the nonlocal form factors
in the vacuum (gravitational) actions were calculated and analysed.
The problem is that, these nonlocal form factors describe the
decoupling, but only for the fourth-derivative terms in the action.
Owing to covariance, the form factors depend on the d'Alembertian
operator $\cx$. The positive powers of this operator give zero when
acting on the cosmological constant and produce surface terms when
acting on the scalar curvature $R$. Let us note that part of the
mentioned papers, Refs.~\cite{Codello,Omar-FF4D}, include the
discussion of the
form factors of surface terms (see \cite{Sebastian} for the latest
discussions of the mathematical aspects of the problem), and
there may be even interesting applications of the running of Newton
constant, related to these surface terms. However, it is unclear how
one can gain information about the running of the cosmological
constant in the traditional covariant framework.

The situation changes dramatically if we perform a conformal
transformation. For instance, using  the parametrization
\beq
g_{\mu\nu} \,=\, \frac{\phi^2}{M^2}\,\exp \big\{\bar{h}_{\mu\nu}
\big\},
\qquad
\exp \big\{\bar{h}_{\mu\nu} \big\}
\,=\, \eta_{\mu\nu} + \bar{h}_{\mu\nu}
+ \frac12\,\bar{h}_{\mu\la}\bar{h}^\la_{\,\,\nu}
+ \frac{1}{3!}\,\bar{h}_{\mu\la}\bar{h}^\la_{\,\,\tau}
\bar{h}^\tau_{\,\,\nu}
+ \dots\,\,,
\label{metpar}
\eeq
with the traceless $\bar{h}_{\mu\nu}$  and constant scale
parameter $M$, transforms the cosmological constant term
$\sqrt{-g}$ into $\phi^4$ plus $\phi^4 \bar{h}^n$-vertices. It
is known that there is no problem to find nonlocal form factor
and verify IR decoupling for the  $\phi^4$-term in the scalar
theory \cite{bexi} and one should expect this to be equally
easy in the gravitational version of the theory.

Unfortunately, the described approach does not constitute a
comprehensive solution of the problem of the running cosmological
constant. In particular, it is not obvious
that such a non-covariant running will preserve the structure of
the   $\phi^4 \bar{h}^n$-vertices, such that the running can be
safely attributed to the cosmological constant and not to the
artificial scheme of reparametrization. Anyway, the running of
the cosmological term in the $4D$ (four-dimensional) theory of
the conformal factor of
the metric is an attractive object of study, starting from the first
proposal \cite{OdSh-91} and its realization by Antoniadis and
Mottola \cite{antmot}. The model of quantum conformal factor
follows the idea to perform secondary quantization of the anomaly
induced effective action of vacuum. This action appears as a result
of integrating conformal anomaly \cite{rie,frts84} coming from the
quantum effects of matter fields (see, e.g., 
\cite{duff94} for the
review or \cite{OUP} for the textbook level introduction). The
simplest realization of the anomaly induced action is a theory of a
single scalar field with fourth derivatives, on a flat background.
This procedure corresponds ``switching off'' the
$\bar{h}_{\mu\nu}$-mode in the parametrization (\ref{metpar}).

In the paper \cite{antmot} it was shown that such a model, with
additional Einstein-Hilbert and cosmological terms, is renormalizable
and, in particular, describes the running of the cosmological term.
The remaining question is whether this running holds in the low
energy domain or only in the UV. Indeed, this is a general question
that is quite relevant for all higher derivative models of quantum
gravity. These models may be renormalizable \cite{Stelle77}, or even
superrenormalizable \cite{highderi} and this enables one to
consistently derive the renormalization group equations for the
effective charges. In the $4D$ 
case, the beta functions are
partially ambiguous \cite{frts82,avbar86,a}, while in the six- or
higher-derivative models, all beta functions do not depend on the
gauge fixing conditions \cite{SRQG-betas}. However, in which
physical situations the corresponding running can be applied? The
one-loop corrections behind the beta functions come from the
three different types of diagrams: \ \textit{(i)} with internal lines
of the massless degrees of freedom (gravitons);  \ \textit{(ii)} with
internal lines of massive components, i.e., higher derivative ghosts
(or ghost-like states, ghost tachyons, etc)
and normal degrees of freedom, typical for the superrenormalizable
models;  \ \textit{(iii)} with mixed (massless and massive) internal
lines.
The standard approach to effective quantum gravity
\cite{Don-94} assumes that only the first and third types
of diagrams give relevant contributions in the IR and that these
contributions are the same as in the effective model where the
propagators and vertices are constructed from the action of GR.
As a consequence, the IR limit of an arbitrary model of quantum
gravity corresponds to the quantum GR, which does not have
massive degrees of freedom. As with all reasonable assumptions,
this statement has to be verified. A relevant question, posed in
\cite{Gauss} and, in a more explicit form, in \cite{Polemic}, was
whether it is possible that, instead of quantum GR, the IR theory of
quantum gravity may be based on some other theory, e.g., based on
a nonlocal action, such that there are still no massive  degrees of
freedom propagating in the IR. Answering this question requires
making one-loop calculation in the momentum-dependent scheme
of renormalization, in a theory with higher derivatives. Such
calculation is very complicated and it looks reasonable to consider
a toy model which possesses the same main features (i.e., massive
and massless particle contents owing to higher derivatives and
non-polynomial interactions). This kind of a model would enable
one to perform necessary calculation in a more economic way.

It is easy to note that the theory of quantum conformal factor
\cite{antmot} represents a nearly perfect toy model for the
fourth-derivative quantum gravity. The Lagrangian of this
theory includes non-polynomial interactions in the two-derivative
and zero-derivative sectors, similar to the fourth derivative quantum
gravity. This means, the general structure of the relevant diagrams
includes all the aforementioned \textit{(i)},  \textit{(ii)} and
\textit{(iii)}-types. Regardless the calculation of the form factors
in the momentum subtraction scheme in the theory  \cite{antmot}
are rather involved (as the reader may see in what follows), they
are still alleviated compared to the ones in a full version of
quantum gravity, where one has to face more extensive set of
degrees of freedom and complicated tensor structures, typical for
diagrammatic treatment of quantum gravity.

In the present work, we report on the derivation of nonlocal
form factors in the fourth-derivative model of quantum conformal
factor and perform the analysis of the UV and IR asymptotic
behaviour of  these quantum corrections.
It is important to note
that the effective approach to the theory of conformal factor
induced by anomaly has an independent interest. In the recent
paper \cite{Mottola-2017}, it was shown that this theory provides,
in the effective approach, a propagation of a scalar mode of the
gravitational field, which is not present in GR. In our opinion,
the investigation of quantum IR decoupling is useful for a better
general understanding of this model in the effective framework.

The paper is organized as follows: In sect.~\ref{sec2}, we briefly
review the four-derivative model for the conformal factor and
present the derivation of its UV divergences using the heat-kernel
method. The corresponding expression will be used, in what
follows, as a reference to verify the main result in the UV.
In sect.~\ref{sec3}, we formulate the elements of Feynman
technique, i.e., the propagator and vertices for the model, and
consider the diagrams producing ultraviolet (UV) divergences.
Furthermore, we derive the one-loop corrections, including the
nonlocal form factors in the propagator sector. Section \ref{sec4}
includes a description of the asymptotic behavior of nonlocal
contributions to the two-point function in the UV and IR limits.
In sect.~\ref{sec5} we discuss the connection between the momentum
dependence in the IR regime of the fundamental theory and the
divergences in the effective low-energy model containing only the
light (massless) mode. As usual in massless theories, the divergences
define not only UV, but also the IR behaviour of the theory and can
be used for comparison with the IR limit of the full theory. In
Sec.~\ref{secCC} we present a discussion of the implications of
the IR decoupling for the cosmological constant problems.
Finally, in Sect.~\ref{sec6}, we draw our conclusions and discuss
the possibilities of a subsequent work.
The four Appendices complement the main text. In the Appendix
\ref{secA}, one can find the set of the Feynman diagrams used in
our calculations, while in Appendix \ref{secB}, we collect
intermediate formulas concerning the calculation of Feynman
integrals in dimensional regularization.
Contributions to the two-point function from divergence-free diagrams
are shown in Appendix \ref{secC}, and in Appendix \ref{secD} we present
the complete expressions of the one-loop quantum corrections to the
three- and four-point vertices. 

The notations include the Minkowski signature $ (+,-,-,-) $. Also, to
reduce the size of the formulas, we avoid indicating the $+i\ep$ in the
denominators of the propagators. Indeed, the loop calculations were
performed in Euclidean signature.

\section{The model} 
\label{sec2}

Let us start with a brief review of the model which we shall use
in the present work. The action of the model is the simplest form
of the solution of the anomaly-induced action \cite{rie,frts84}, with
the flat fiducial metric, plus the Einstein-Hilbert and cosmological
terms,
\beq
\label{eff_action}
S_{\textrm{cf}}
\,=\,
\int d^{4}x\left\{
2b(\square\si)^{2}-\left(2w+2b+3c\right)
\left[\square\si+(\pa\si)^{2}\right]^2
+\frac{3}{\ka}e^{2\si}(\pa\si)^{2}-\frac{\La}{\ka}e^{4\si}\right\}.
\quad
\eeq
Here $\ka=8\pi G$ and the coefficients $w$, $b$ and $c$ are the
one-loop semiclassical beta functions in the vacuum sector,
\beq
&&
w=\frac{1}{120(4\pi)^{2}}(N_{s}+6N_{f}+12N_{v}),
\nn
\\
&&
b=-\frac{1}{360(4\pi)^{2}}(N_{s}+11N_{f}+62N_{v}),
\nn
\\
&&
c = \frac{1}{180(4\pi)^2}(N_{s}+6N_{f}
-18N_{v}),
\label{wbc}
\eeq
where $N_{s},\,N_{f},\,N_{v}$ are the multiplicities of the
quantum conformal matter fields of spins zero, one-half and
one, respectively. The trace anomaly which produces the
induced part of the action (\ref{eff_action})  is
\beq
\langle T^\mu_{\,\,\mu} \rangle \,=\,-\big( wC^2 + bE_4
+ c\,{\square}R \big) .
\label{T-II}
\eeq
The coefficient $c$ can be modified by adding a finite local term
$R^2$ to the action $S_{anom}$ (see
\cite{duff94,anomaly-2004,BoxAno,AnInt22} for detailed discussion).
This feature will not affect our considerations, especially because
we will not need particular versions of the beta functions
(\ref{wbc}) and concentrate on the general features of the quantum
theory of conformal factor based on (\ref{eff_action}).

On top of induced part, the action includes Einstein-Hilbert
and cosmological terms, which are not renormalized at the
initial semiclassical theory, but become very relevant at the
second stage, when we quantize the conformal factor.

 The idea that the conformal factor can be quantum, despite it
 emerges as an effective mode in the integration of matter fields,
 comes from Polyakov's approach in $2D$, related to string
 theory \cite{Poly81}. The idea of using the equivalent metric-scalar
 (Liouville) model as the basis of $2D$ quantum gravity was quite
 popular in 90-s.  The use of the analogous theory (in curved
 spacetime) as a model for $4D$ quantum gravity was proposed
 in \cite{OdSh-91}. In four dimensions, the theory for the
conformal factor can be regarded as a truncated version of the
four-derivative quantum gravity at large distances (i.e., for the
low energies, or IR), providing a screening mechanism for the
cosmological constant \cite{antmot}. An important difference
with the  $2D$ induced gravity is that, in  $4D$ one can add the
classical terms. Alternatively, one can make the Einstein-Hilbert
and cosmological terms to be generated in the scheme of induced
gravity \cite{induce}, but this requires an independent scalar field
and does not fit our purpose to construct a simplified model to
explore the decoupling in a higher derivative quantum gravity.

As any fourth-derivative  quantum gravity model, the model
of our interest has massive modes, which can be ghosts and
tachyonic ghosts.\footnote{
It is worth mentioning that in
Ref.~\cite{AntMazMot97} it was argued that ghosts are
eliminated in the pure anomaly theory, which is equivalent
to (\ref{eff_action}), by imposition of the constraints of
diffeomorphism invariance. This may be an indication that
in any theory of gravity, diffeomorphism invariance and
the constraints they impose must play a role.}
The question of our interest is what happens
with the contributions of these massive modes at low energies.

It proves useful to introduce notations similar  to \cite{antmot},
\beq
\label{parameters2}
\theta^{2} \equiv (2w+3c),\;\;\;\ze\equiv \big(2w+2b+3c\big),
\qquad
\ga\equiv\dfrac{3}{\ka}\,,
\qquad
\textrm{and}
\qquad
\la\equiv\dfrac{\La}{\ka},
\eeq
such that the action \eqref{eff_action} becomes
\beq
\label{eff_action2}
S_{\textrm{cf}} & = & \int d^{4}x\,
\Bigl\{ -\th^{2}(\square\si)^{2}
- \ze\left[2(\pa\si)^{2}\square\si
+ (\pa\si)^{4}\right]
+\ga\,e^{2\si}(\pa\si)^{2}-\la\,e^{4\si}\Bigr\}. \;\;\;\;\;
\eeq
The difference in notations with the paper by Antoniadis and Mottola
is the coefficient of the kinetic sector of higher derivative terms
$\th^2$, which is denoted $Q^{2}/(4\pi)^{2}$ in \cite{antmot}. On
top of this, we fix the anomalous scaling dimension $\al=1$ to avoid
additional complications of formulas. 

The last two terms in \eqref{eff_action2} come from the
Einstein-Hilbert and cosmological constant terms. In the IR,
these terms dominate over the higher derivative terms and
it proves useful to split the Lagrangian into two terms, i.e.,
\beq
\label{partIR_S}
\mathcal{L}_{\textrm{IR}}=\ga e^{2\si}(\pa\si)^{2}-\la e^{4\si}
\eeq
and
\beq
\label{part4_S}
\mathcal{L}_{4\textrm{der}}
= -\th^{2}(\square\si)^{2}
- \ze\left[2(\pa\si)^{2}\square\si
+ (\pa\si)^{4}\right].
\eeq

Our plan is to evaluate the quantum corrections in full theory
(\ref{eff_action2}) and, separately, for the theory based on the
IR-term (\ref{partIR_S}). Due to the presence of higher derivative
terms, the one-loop divergences in the full theory are obtained
using the generalized Schwinger-DeWitt technique
\cite{frts82,bavi85}.

Using the background field method, the conformal factor is
decomposed into classical $\si$ and quantum $\rho$ counterparts,
$ \si\rightarrow\si+\rho $. Then we obtain the bilinear in the
quantum field forms for the two terms,
\beq
\label{part4_bi}
\mathcal{\mathcal{S}}_{4\textrm{der}}^{(2)}
& = &
-\int d^{4}x\Big\{ \th^{2}(\square\rh)^{2}
+ 2\ze\big[(\pa\rh)^{2}\square\si
+ 2(\pa_{\mu}\rh)(\pa^{\mu}\si)\square\rh
+(\pa\si)^{2}(\pa\rh)^{2}
\nn
\\
&&
+\,\,
2(\pa_{\mu}\rh)(\pa_{\nu}\rh)(\pa^{\mu}\si)(\pa^{\nu}\si)\big]\Big\}
\eeq
and
\beq
\label{partIR_bi}
\mathcal{\mathcal{S}}_{IR}^{(2)}=\int d^{4}x\Big\{\ga e^{2\si}\big[(\pa\rh)^{2}+4\rh(\pa_{\mu}\rh)(\pa^{\mu}\si)
+2\rh^{2}(\pa\si)^{2}\big]-8\la\rh^{2}e^{4\si}\Big\}.
\eeq
The Hermitian forms for the structures \eqref{part4_bi} and
\eqref{partIR_bi} are obtained as
\beq
\frac{\delta^{2}
\mathcal{\mathcal{S}}_{4\textrm{der}}^{(2)}}{\de\rh(y)\de\rh(z)}
& = &
-\,2\th^{2}\square^{2}+4\ze\big[2(\square\si)\square
-2\pa^{\mu}(\pa^{\nu}\si)\pa_{\mu}\pa_{\nu}
+4(\pa_{\nu}\si)\pa^{\mu}(\pa^{\nu}\si)\pa_{\mu}
\nn
\\
&&
+\,\,
(\pa\si)^{2}\square+2(\square\si)(\pa^{\mu}\si)\pa_{\mu}
+2(\pa^{\mu}\si)(\pa^{\nu}\si)\pa_{\mu}\pa_{\nu}\big],
\nn 
\\
\frac{\de^{2}\mathcal{\mathcal{S}}_{IR}^{(2)}}{\de\rh(y)\de\rh(z)}
& = &
-\,\,2\ga e^{2\si}\big[\square+2(\pa^{\mu}\si)\pa_{\mu} + 2(\pa\si)^2
+2\square\si\big]-16\la e^{4\si}.
\label{Herm_SIR}
\eeq
So, for the complete model \eqref{eff_action2}, we have
\beq
\frac{\de^{2}\mathcal{\mathcal{S}}^{(2)}}{\de\rh(y)\de\rh(z)}
= -2\th^{2}\hat{H},
\label{Herm_S}
\eeq
where the self-adjoint four-derivative minimal operator is
\beq
\hat{H} = \square^2
+ V^{\mu\nu}\pa_{\mu}\pa_{\nu}
+ N^\mu \pa_\mu
+ U,
\eeq
with the elements
\beq
&&
V^{\mu\nu} = -\frac{2\ze}{\th^{2}}\big[2\eta^{\mu\nu}\square\si
- 2\pa^{\mu}\pa^{\nu}\si+\eta^{\mu\nu}(\pa\si)^{2}
+ 2(\pa^{\mu}\si)\pa^{\nu}\si\big]
+ \frac{\ga}{\th^{2}} e^{2\si}\eta^{\mu\nu},
\nn
\\
&&
N^{\mu} = -\frac{4\ze}{\th^{2}}
\big[2(\pa_{\nu}\si)\pa^{\mu}\pa^{\nu}\si
+(\square\si)\pa^{\mu}\si\big]
+\frac{2\ga}{\th^{2}}e^{2\si}(\pa^{\mu}\si),
\nn
\\
&& U = \frac{2\ga}{\th^{2}}e^{2\si}\big[(\pa\si)^{2}+\square\si\big]
+\frac{8\la}{\th^{2}}e^{4\si}.
\label{VU}
\eeq
Using the standard algorithm for the fourth-order operators
\cite{frts82,bavi85}, we arrive at the expression for the divergences
\beq
\label{div_4}
\bar{\Ga}_{\textrm{div}}^{(1)} = -\frac{1}{\vp}\int d^{4}x\left\{\frac{5\ze^{2}}{\th^{4}}\left[\square\si
+ (\pa\si)^{2}\right]^{2}
+\frac{\ga}{\th^{2}}\Big(\frac{3\ze}{\th^{2}}
+2\Big)(\pa\si)^{2}e^{2\si}
- \Big(\frac{8\la}{\th^2}
- \frac{\ga^{2}}{2\th^4}\Big) e^{4\si}\right\},
\eeq
where we introduce the useful notation $\vp=(4\pi)^{2}(n-4)$ and
neglect the irrelevant surface terms. This result agrees with the 
previous calculations \cite{antmot,AntoOdin94}, except for an
apparent misprint in the sign of Eq.~(4) of \cite{AntoOdin94}. An 
additional verification can be found in the recent paper \cite{Holdom}.

In that follows, we shall confirm the expression
(\ref{div_4}) by the calculation of both divergent and finite
nonlocal (leading logarithms) parts of the Feynman diagrams.
By considering the minimal subtraction (MS) scheme, one can
easily derive UV  $\be$-functions for the theory \eqref{eff_action2}.
In the next section, we will determine the finite parts of the one-loop
diagrams that produce these divergences. In this case, the structure
\eqref{div_4} will be useful in identifying the diagrams that are
relevant for our purposes.
For example, from the coefficients of the terms $(\pa\si)^{2}e^{2\si}$
and $e^{4\si}$, we can expect that diagrams
involving interaction vertices $\ga$ and $\ga\ze$ provide contributions
to the Einstein-Hilbert sector, while diagrams with vertices $\la$ and
$\ga^{2}$ are associated with corrections to the cosmological
constant sector. Later on, we shall see that this identification may
not hold in the effective low-energy model with two derivatives. 

For completeness, we also derived the divergences of the
effective theory, based on the IR-term, Eq.~\eqref{partIR_S},
separately. The result is
\beq
\label{div_IR}
\bar{\Ga}_{\textrm{div,\,IR}}^{(1)}=-\frac{1}{\vp}\int
d^{4}x\,\bigg\{\frac{1}{2}\big[\square\si
+(\pa\si)^{2}\big]^{2}
-\frac83\La\,e^{2\si}(\pa\si)^{2}
+\frac{32}{9}\La^{2}e^{4\si}\bigg\}.
\eeq
As it should be expected from the power counting, the
fourth-derivative counterterms are required in this theory,
as it is non-renormalizable. At the same time, neglecting the
fourth-derivative terms according to the effective approach,
we arrive at the reference expression to compare with the
IR limit of the full theory.

\section{One-loop corrections from Feynman diagrams}
\label{sec3}

In a model with higher derivatives, to explore the decoupling in the
loop corrections, one has to separate massive and massless degrees
of freedom. In many cases, this can be achieved by introducing
auxiliary fields (see, e.g., \cite{Cremin}).  However, in the case of
the theory \eqref{eff_action2}, this approach is not operational
owing to our interest in the quantum corrections in the theory that
have higher derivatives in both kinetic terms and the interactions.
Thus, we shall make the separation at the level of the propagator
and vertices in the Feynman diagrams, i.e., use the method close to
the one of \cite{julton}.

The structure of the vertices and the propagator for the fundamental
theory \eqref{eff_action2} can be calculated by using the
parametrization $\si \rightarrow \si+\rh$, where $\rh$ is a small
perturbation and expanding the exponential terms in the power
series in $\rh$. Collecting the quadratic terms, we find that the
propagator satisfies the equation
\beq
\label{Greenf}
2\big[\th^{2}\square^{2} + \ga\square + 8\la\big]
G(x,y)\,=\,i\de^{4}(x-y).
\eeq
Making the Fourier transform,
\beq
G(x,y)=\int \frac{d^{4}k}{(2\pi)^{4}}\,e^{-ik\cdot(x-y)}\widetilde{G}(k)
\label{Furier}
\eeq
and assuming $\La \ll \ga^2/\theta^2$, we get
\beq\label{prop}
\widetilde{G}(k)  \,\,=\,\,
\frac{i}{2[\th^{2}k^{4}-\ga k^{2}+8\la]}
\,\,\simeq\,\,
\frac{i}{2\,\th^2\big(k^{2}-\frac{\ga}{\th^{2}}\big)
\big(k^{2}-\frac{8}{3}\La\big)}.
\eeq
Finally, in the same approximation, the propagator can be written as
\beq
\label{gpro}
\widetilde{G}(k)\,\,=\,\,
\frac{i}{2\,\th^2\big(m^{2}-M^{2})}
\bigg[
\frac{1}{k^2-m^2} - \frac{1}{k^2-M^2}\bigg].
\eeq
It is easy to identify a healthy degree of freedom with the mass
$m^2=8\La/3$ and a ghostly mode with the Planck-scale
mass, $M^2=\ga/\th^2$.

We need to consider only those interaction vertices that are relevant
for the one-loop corrections to the propagator. The vertices for the
$3$- and $4$-point functions arise from the derivative interaction
terms in the part $\mathcal{L}_{4\textrm{der}}$ and from the
higher order terms in the exponential expansion in
$\mathcal{L}_{\textrm{IR}}$,
\beq
\ze\big[2(\pa\rh)^{2}\square\rh+(\pa\rh)^{4}\big],
\qquad
2\ga(\pa\rh)^{2}\big[\rh+\rh^{2}\big], ,
\qquad
\frac{32\la}{3} \big[\rh^{3}+\rh^{4}\big]. 
\eeq

\begin{figure}[H]
\centering
\includegraphics[scale=0.52]{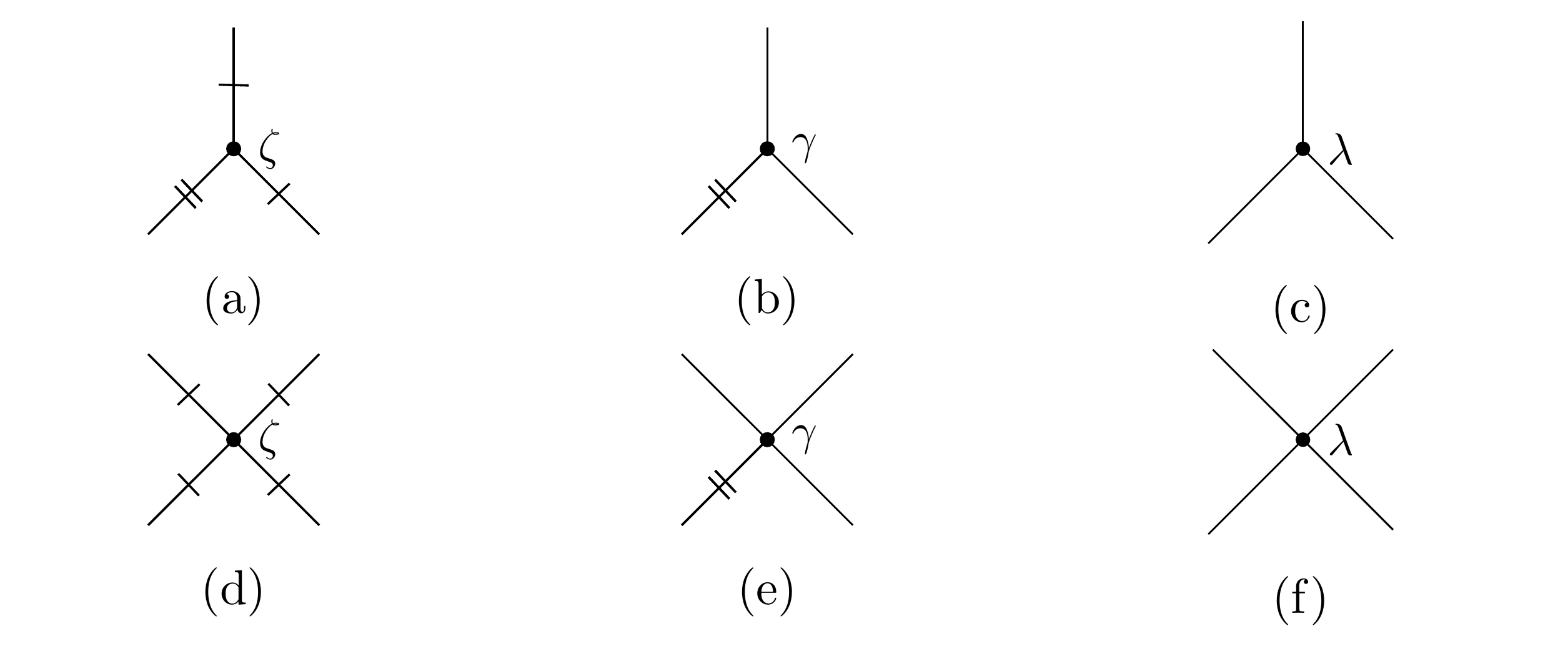}
\begin{quotation}
\vspace{-6mm}
\caption{\small
Feynman diagrams associated with the interaction vertices.
The primes denotes derivatives acting on the propagators.}
\label{fig1}
\end{quotation}
\end{figure}
\vspace{-1cm}

In Fig.~\ref{fig1},
we have presented the vertices corresponding to these interaction terms.
The analytic expressions have the form
\beq
&&
V^{(3)}_{\ze}(p,k,q)= -4i\ze \big[p^{2}(k\cdot q)+q^{2}(k\cdot p)
+k^{2}(p\cdot q)\big], 
\nn
\\
&&
V^{(3)}_{\ga}(p,k,q)= 2i\ga(k^{2}+p^{2}+q^{2}),
\nn
\\
&&
V^{(3)}_{\la}= -64i\la
\label{3la}
\eeq
and
\beq
&&
V^{(4)}_{\ze}(p,k,q,r)
=  -8i\ze \big[(k\cdot q)(p\cdot r)
+ (k\cdot p)(q\cdot r) + (k\cdot r)(p\cdot q)\big],
\nn
\\
&&
V^{(4)}_{\ga}(p,k,q,r)= 4i\ga(k^{2}+p^{2}+q^{2}+r^{2}),
\nn 
\\
&&
V^{(4)}_{\la}= -256i\la,
\label{4la}
\eeq
where $(k\cdot p)=k_\mu p^\mu$.

Now we are in the position to determine the one-loop contributions
to the self-energy (correction to the propagator) of the healthy (light)
mode. First, consider the diagrams producing the UV divergences
\eqref{div_4}, derived previously using the heat-kernel method. These
divergences are responsible for the MS-scheme based beta functions
and serve as the UV references for the complete expressions.

In the theory \eqref{eff_action2} the expression for the
two-point function is
\beq
\label{g2}
G^{(2)}_{\textrm{1-loop}}(p,-p)\propto \frac{i}{(p^{2}-m^{2})}
\big(\bar{\Si}_{\ga}
+ \bar{\Si}_{\la}
+\widetilde{\Si}_{\ga\la}
+\widetilde{\Si}_{\la^{2}}
+\Si_{\ze^{2}}
+\Si_{\ga\ze}
+\Si_{\ga^{2}}
+\ldots\big)\frac{i}{(p^{2}-m^{2})}, \;
\eeq
where the omitted $(\ldots)$ terms denote
contributions from the divergence-free diagrams. On top of this, the finite contributions $\Si_{\la\ze},\;\Si_{\ga\la}$ and $\Si_{\la^{2}}$ are proportional to the mass $m^{2}$ (as shown in Appendix \ref{secC}). As we shall see in Sec.~\ref{sec4}, these terms can be neglected in our analysis of the IR limit, when we consider $m^{2}=0$. 

\begin{figure}[H]
\centering
\includegraphics[scale=0.22]{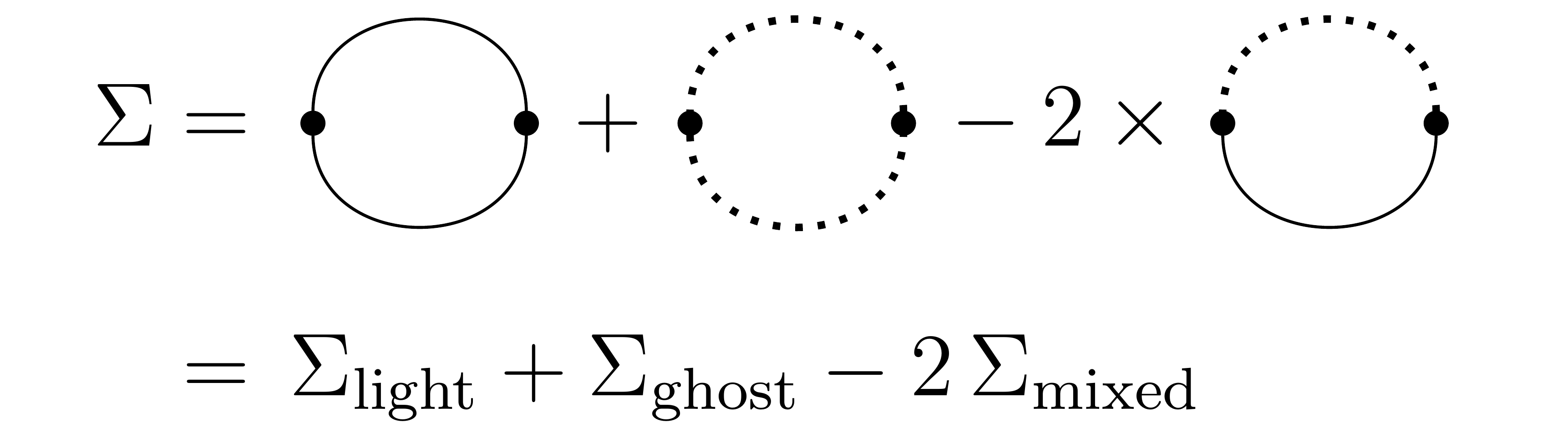}
\begin{quotation}
\vspace{-6mm}
\caption{\small
General structure of one-loop diagrams for self-energy functions $\Si$. Solid lines indicate light degrees of freedom while dashed lines stand for the massive ghosts.
}
\label{self_energy}
\end{quotation}
\end{figure}

\begin{figure}[H]
\centering
\includegraphics[scale=0.25]{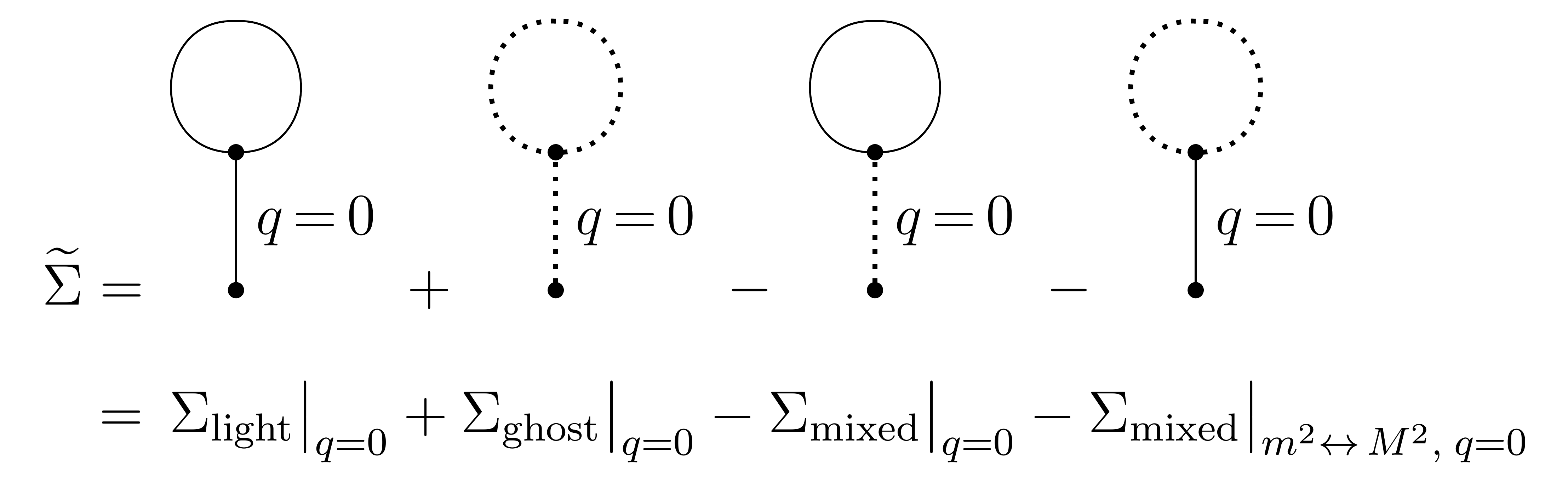}
\begin{quotation}
\vspace{-6mm}
\caption{\small
``Tadpole" diagrams associated with the corrections $\widetilde{\Si}$.}
\label{tadpole_diag}
\end{quotation}
\end{figure}

We can write the general symbolic expression for the self-energy function, in the second-order (in coupling constants)
approximation as presented in Fig.~\ref{self_energy}, for corrections of the type $\Si$, and in Fig.~\ref{tadpole_diag}, for corrections $\widetilde{\Si}$. 
\vspace{-1cm}

\begin{figure}[H]
\centering
\includegraphics[scale=0.145]{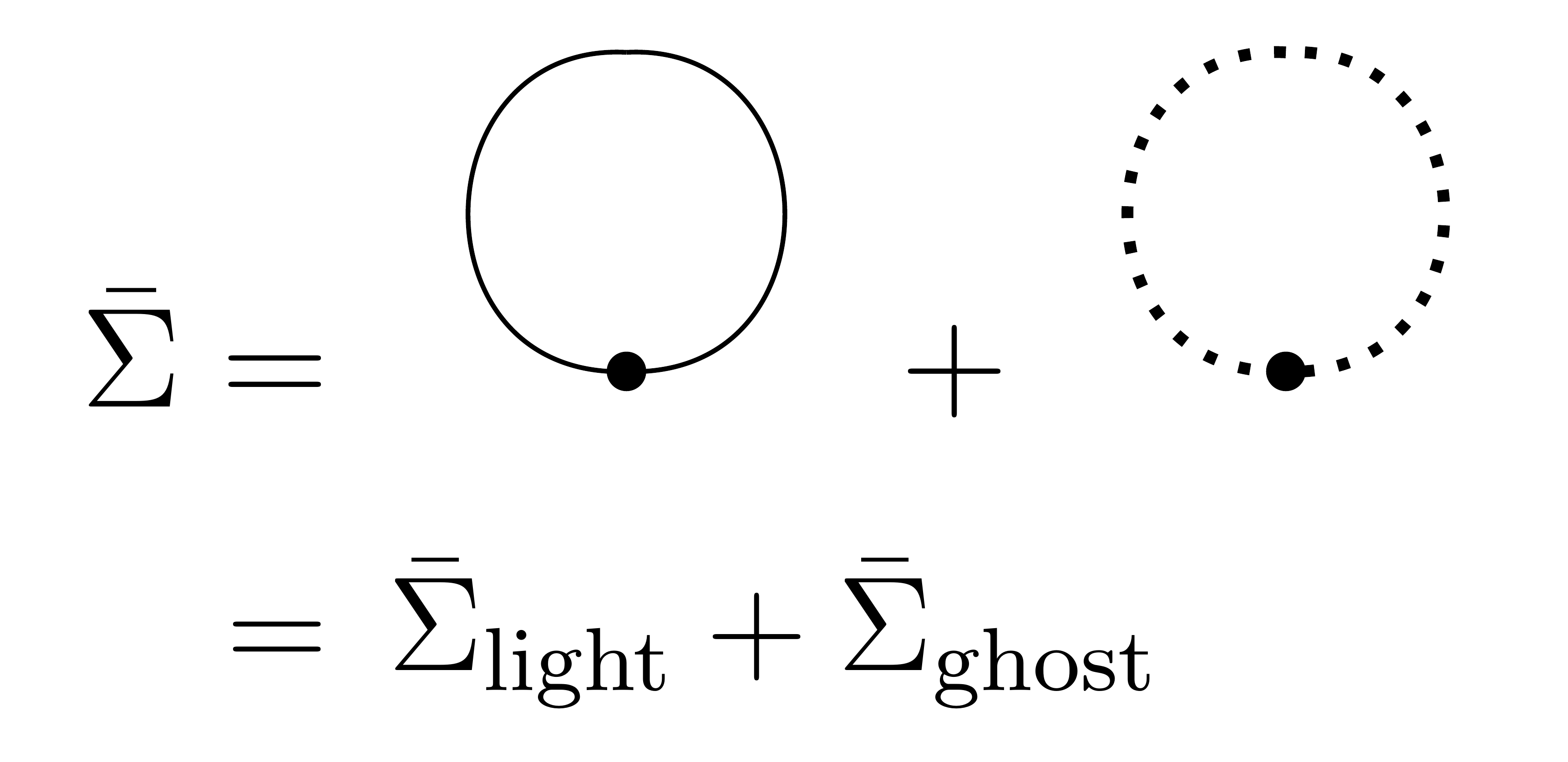}
\begin{quotation}
\vspace{-6mm}
\caption{\small
``Snail" diagrams associated with the corrections $\bar{\Si}$. 
In this case, obviously, there are no contributions from a mixed sector. }
\label{snail_diag}
\end{quotation}
\end{figure}

\vspace{-1cm}
\noindent
Furthermore, in Figs.~\ref{self_energy} and \ref{tadpole_diag}, the 
expressions for the light, ghost and mixed sectors are, respectively,
\beq
&&
\Si_{\textrm{light}} \propto \int \frac{d^{4}k}{(2\pi)^{4}}
\,\frac{i}{(k^{2}-m^{2})}
V^{(3)}(p,-k,k-p)\frac{i}{\big[(k-p)^{2}-m^{2}\big]}V^{(3)}(-p,k,p-k),
\mbox{\qquad}
\nn
\\
&&
\Si_{\textrm{ghost}} \propto \int \frac{d^{4}k}{(2\pi)^{4}}\,\frac{i}{(k^{2}-M^{2})}V^{(3)}(p,-k,k-p)
\frac{i}{\big[(k-p)^{2}-M^{2}\big]}V^{(3)}(-p,k,p-k),
\mbox{\qquad}  
\nn
\\
&&
\Si_{\textrm{mixed}} \propto \int \frac{d^{4}k}{(2\pi)^{4}}
\,\frac{i}{(k^{2}-m^{2})} V^{(3)}(p,-k,k-p)
\frac{i}{\big[(k-p)^{2}-M^{2}\big]}V^{(3)}(-p,k,p-k),
\mbox{\qquad}
\eeq
while the contributions of the first-order in the coupling constants
have the general structure shown in Fig.~\ref{snail_diag}, 
\vspace{-1cm}
\noindent
with
\beq
&& \bar{\Si}_{\textrm{light}}
\propto
\int \frac{d^{4}k}{(2\pi)^{4}}\,
\frac{i}{(k^{2}-m^{2})}V^{(4)}(p,k,-p,-k),
\nn
\\
&&
\bar{\Si}_{\textrm{ghost}} \propto
\int \frac{d^{4}k}{(2\pi)^{4}}\,\frac{i}{(k^{2}-M^{2})}
V^{(4)}(p,k,-p,-k).
\eeq

The diagrammatic representation of the contributions to the
two-point function with different couplings, is presented in the
Appendix \ref{secA}. The diagrams in Fig.\,\ref{fig2} correspond
to the fifth term in \eqref{g2}, and the last term is associated with
the diagrams in Fig.\,\ref{fig4}. In Fig.\,\ref{fig3} there are shown
the diagrams that correspond to the term proportional to $\ga\ze$.
The diagrams for the first-order terms in the couplings $\ga$ and
$\la$, are depicted in Figs.~\ref{fig5} and \ref{fig6}, respectively.
In addition, the third and fourth terms  in \eqref{g2} are associated
with the tadpole diagrams with interaction vertices $ \ga\la $
shown in Fig.\,\ref{fig7}, and $\la^{2}$ shown in Fig.\,\ref{fig8}.

Let us note that each diagram here represents the sum over all the
topologically equivalent diagrams with different permutations over
the external momenta and with all possible placements of derivatives
on the internal and external lines. On top of that, we omitted some
tadpole-type diagrams that do not contribute to $G^{(2)}(p,-p)$,
as they include derivatives of the propagator in a single spacetime
point, and hence vanish.

To evaluate the integrals in \eqref{g2} we used dimensional
regularization. In the model under consideration, this requires
extending the standard list of divergent expressions \cite{Leibb75}
for the integrals in the spacetime of $2\om$ complex dimensions.
The integrals proportional to $\ze^{2}$, $\ga\ze$ and $\ga^2$ read,
respectively, as
\beq
\label{DG_ze2}
\Si^{(2\om)}_{\ze^{2}}(p)
& = &
-\,\frac{8\ze^{2}}{\th^{4}(m^{2}-M^{2})^{2}}
\int \frac{d^{2\om}k}{(2\pi)^{2\om}}\,\Ga^{(3,3)}_{\ze^{2}}
\biggl\{\frac{2}{(k^{2}-m^{2})\big[(k-p)^{2}-M^{2}\big]}
\nn
\\
&& -\,\,
\frac{1}{(k^{2}-m^{2})\big[(k-p)^{2}-m^{2}\big]}
\,-\,\frac{1}{(k^{2}-M^{2})\big[(k-p)^{2}-M^{2}\big]}
\biggr\},
\eeq
\beq
\label{DG_gaze}
\Si^{(2\om)}_{\ga\ze}(p)
& = &
-\,\frac{4\ga\ze}{\th^{4}(m^{2}-M^{2})^{2}}\int \frac{d^{2\om}k}{(2\pi)^{2\om}}\,\Ga^{(3,3)}_{\ga\ze}
\biggl\{\frac{2}{(k^{2}-m^{2})\big[(k-p)^{2}-M^{2}\big]}
\nn
\\
&& -\,\,\frac{1}{(k^{2}-m^{2})\big[(k-p)^{2}-m^{2}\big]}
\,-\,\frac{1}{(k^{2}-M^{2})\big[(k-p)^{2}-M^{2}\big]}\biggr\}
\eeq
and
\beq
\label{DG_ga2}
\Si^{(2\om)}_{\ga^{2}}(p)
& = &
-\,\frac{2\ga^{2}}{\th^{4}(m^{2}-M^{2})^{2}}\int \frac{d^{2\om}k}{(2\pi)^{2\om}}\,\Ga^{(3,3)}_{\ga^{2}}
\biggl\{\frac{2}{(k^{2}-m^{2})\big[(k-p)^{2}-M^{2}\big]}
\nn
\\
&& -\,\,\frac{1}{(k^{2}-m^{2})\big[(k-p)^{2}-m^{2}\big]}
-\frac{1}{(k^{2}-M^{2})\big[(k-p)^{2}-M^{2}\big]}\biggr\},
\eeq
where we used the following combinations of the vertex factors:
\beq
&&
\Ga^{(3,3)}_{\ze^{2}} \,=\,
p^{4}k^{4}-2p^{2}k^{2}(p\cdot k)^{2}+(p\cdot k)^{4},
\nn
\\
&&
\Ga^{(3,3)}_{\ga\ze}\,=\,
p^{4}k^{2}-(p^{2}+k^{2})(p\cdot k)^{2}
+(p\cdot k)^{3}-p^{2}k^{2}(p\cdot k)+k^{4}p^{2},
\nn
\\
&&
\Ga^{(3,3)}_{\ga^{2}} \,=\,
p^{4}+k^{4}+(p\cdot k)^{2}+2p^{2}k^{2}-2(p^{2}+k^{2})(p\cdot k).
\label{vf33}
\eeq
The results of the integrations in the Euclidean space
are\footnote{Some intermediate details of the calculations
can be found in Appendix \ref{secB1}. The calculations
were verified using
the Package-$X$ \cite{Patel} in Mathematica
\cite{Wolfram}.}
\beq
\label{rDG_ze2}  
\Si_{\ze^{2}}(p)
& = &
\frac{i\ze^{2}p^{4}}{(4\pi)^{2}\th^{4}}\,
\bigg\{5\bigg[\frac{1}{\ep}
+\ln\Big(\frac{\mu^{2}}{m^{2}}\Big)\bigg]
-\frac{1}{4}\big[9A^{2}-5(ab)^{2}-37\big]
-\frac{1}{2(ab)^{2}c^{5}}\ln\Big(\frac{1+c}{1-c}\Big)
\nn
\\
&&
- \,\frac{1}{2(ab)^{2}d^{5}}\ln \Big(\frac{1+d}{1-d}\Big)
- \bigg[\frac{1}{2}(ab)^{3}
+\frac{5}{2}ab \Big(ab+\frac{a}{2}+2\Big)
+\frac{15a}{4}\Big(1+\frac{1}{4b}\Big)
\nn
\\
&&
+ \,
5\Big(2+\frac{3}{4b}+\frac{1}{2ab}\Big)\bigg] \ln(1+4b)
+\frac{A^{5}}{2(ab)^{2}}
\ln \Big[\frac{(A+1)^{2}-(ab)^{2}}{(A-1)^{2}-(ab)^{2}}\Big]\bigg\},
\eeq
\beq
\label{rDG_gaze}
\Si_{\ga\ze}(p)
& = &
- \, \frac{i\ga\ze p^{2}}{(4\pi)^{2}\th^{4}}\,\bigg\{ 3\bigg[\frac{1}{\ep} + \ln\Big(\frac{\mu^2}{m^2}\Big)\bigg]
+ \frac{1}{2c(ab)^{2}}\bigg[\frac{a}{2}
\Big(2 - \frac{1}{c^{2}}\Big) + 1 \bigg]
\ln \Big(\frac{1+c}{1-c}\Big)
\nn
\\
&&
+\,\frac{1}{2d(ab)^{2}}\bigg[\frac{a}{2}
(4b+1)\Big(2 - \frac{1}{d^{2}}\Big) + 1\bigg]
\ln \Big(\frac{1+d}{1-d}\Big) - \Big(ab+\frac{a}{2}-6 \Big)
\nn
\\
&&
- \,\frac{A}{2}\bigg[\Big(\frac{1}{ab} - ab\Big)
\Big(1+\frac{1}{2b}+\frac{1}{ab}\Big)
- \frac{2}{b} \Big(1+\frac{1}{4b}\Big)\bigg]
\ln\Big[\frac{(A+1)^{2}-(ab)^{2}}{(A-1)^{2}-(ab)^{2}}\Big]
\nn
\\
&&
-\, \bigg[\frac{ab}{2}\Big(ab + \frac{a}{2} + 2\Big)
+ 3\Big(1+\frac{a}{2}\Big) \Big(1+\frac{1}{4b}\Big)
- \frac{1}{ab}\bigg] \ln(1+4b) \bigg\},
\eeq
\beq
\label{rDG_ga2}
\Si_{\ga^{2}}(p)
& = & \frac{i\ga^{2}}{(4\pi)^{2}\theta^{4}}\,
\bigg\{
2\bigg[\frac{1}{\ep}+\ln\Big(\frac{\mu^{2}}{m^{2}}\Big)\bigg] + 3
- \frac{1}{2c(ab)^{2}}\Big(\frac{a^{2}}{4}-\frac{1}{c^{2}}+2\Big)
\ln\Big(\frac{1+c}{1-c}\Big)
\nn
\\
&&
+\, \frac{A}{2(ab)^{2}}\Big(ab+\frac{a}{2}-1\Big)^2
\ln \Big[\frac{(A+1)^{2}-(ab)^{2}}{(A-1)^{2}-(ab)^{2}}\Big]
- \frac12\bigg[ab + a\Big(1+\frac{1}{4b}\Big) + 3
\nn
\\
&&
+ \,\frac{1}{2b} \Big(1-\frac{2}{a}\Big) \bigg]\ln(1+4b)
- \frac{1}{2d(ab)^{2}}
\bigg[\frac{a^{2}}{4}(4b+1)^{2}-\frac{1}{d^{2}}+2\bigg]
\ln \Big(\frac{1+d}{1-d}\Big)
\bigg\}, \hspace{.8cm}
\eeq
where
\beq
\dfrac{1}{\ep}\equiv\dfrac{1}{2-\om}-\ga_{\textrm{E}}+\ln(4\pi),
\qquad
a=\dfrac{4m^{2}}{p^{2}},
\qquad
b=\dfrac{M^{2}-m^{2}}{4m^{2}}
\label{epsil}
\eeq
and $\ga_{\textrm{E}} \approx 0.577$ is the Euler-Mascheroni
constant. In the limit $\om\rightarrow 2$, the results
\eqref{rDG_ze2}, \eqref{rDG_gaze} and \eqref{rDG_ga2}
represent divergent and finite parts. Note that the finite part of
these expressions has a very complicated dependence on the
external momentum. For these nonlocal structures, in the mixed
sector, we used the notation
\beq
\label{AA}
A \,=\, \sqrt{(1+ab)^{2}+a} .
\eeq
Furthermore, the notations used in the light and ghost sectors,
include, respectively,
\beq
\label{cd}
c^2  = \dfrac{p^{2}}{p^{2}+4m^{2}},
\qquad
d^2  =  \dfrac{p^{2}}{p^{2}+4M^{2}}.
\eeq
It is worth explaining how the definitions
\eqref{AA} and \eqref{cd} can be used in identifying
contributions coming from the different sectors. E.g., the
logarithmic form factors involving $c$, in the results presented
above, come from the ``pure'' loops with only the propagators
of the light degrees of freedom 
(equivalently, for the other momentum-dependent logarithmic structures).
Of course, there are terms that result from the combination
of contributions from the three sectors (i.e., light, massive ghost
and mixed), such as $\ln(1+4b)$ and those with only polynomial
dependencies on the external momentum.

In top of this, the relevant corrections involving the quartic vertices,
are given by
\beq
\bar{\Si}_{\ga}(p) \,=\,
\frac{2i\ga
p^{2}}{\th^{2}}\,I_{\textrm{quad}}
\qquad
\mbox{and}
\qquad
\bar{\Si}_{\la}(p) = \frac{64i\la }{\th^{2}}\,I_{\textrm{quad}},
\label{1la}
\eeq
where the integral is
\beq
I_{\textrm{quad}}
& = &
\frac{1}{(m^{2}-M^{2})}\int \frac{d^{2\om}k}{(2\pi)^{2\om}}
\,\,\biggl\{\frac{1}{(k^{2}+m^{2})}
-\frac{1}{(k^{2}+M^{2})}\biggr\}
\nn
\\
& = &
-\,\frac{1}{(4\pi)^{2}}\,
\bigg[\frac{1}{\ep}+\ln\Big(\frac{\mu^{2}}{m^{2}}\Big)
\,+\, 1 \,-\,\frac{M^{2}}{(m^{2}
- M^{2})}\ln\Big(\frac{m^{2}}{M^{2}}\Big)\bigg].
\eeq
The results \eqref{1la} are tadpole-type contributions, which
do not produce a nonlocal form factor. Therefore, these
corrections are not relevant to our analysis at low energies
and were included just of completeness.
Of course, the same consideration applies to second-order
tadpole-type corrections, which in principle contribute to divergences
in the Einstein-Hilbert and cosmological constant sectors,
\beq
\widetilde{\Si}_{\ga\la}(p) \,=\,
-\frac{16i\ga\la
p^{2}}{\th^{4}m^{2}M^{2}}\,I_{\textrm{quad}}
\qquad
\mbox{and}
\qquad
\widetilde{\Si}_{\la^{2}}(p)
= -\frac{512i\la^{2} }{\th^{4}m^{2}M^{2}}\,I_{\textrm{quad}},
\label{1gala}
\eeq
with $m^{2}M^{2}=8\la/\th^{2}$.
Actually, the corrections in \eqref{1gala} can be disregarded because tadpole diagrams, such as those presented in Figs.\,\ref{fig7} and \ref{fig8}, normally are eliminated using renormalization conditions (see, e.g., \cite{Collins} or the Chapter 11 of \cite{PeskSchr} for more details).

Let us note that here we presented only the results for the
self-energy. The lower-order vertices were also derived and
produce qualitatively the same picture. The nonlocal parts of
these contributions follow a standard logarithmic structure, as
those for propagator corrections. Since the corresponding
formulas are relatively bulky, they are separated in Appendix
\ref{secD}.

\section{Asymptotic behavior}
\label{sec4}

In this section, we explore the asymptotic behavior of the one-loop
contributions \eqref{rDG_ze2}, \eqref{rDG_gaze}, and \eqref{rDG_ga2}.
Our main interest is to verify how these expressions interpolate
between the UV and IR regions of the fundamental theory. In this
way, we have a chance to understand what happens to the nonlocal
form factors of the contribution of loops with the massive degrees
of freedom (massive ghosts) in the IR.

We start with the limit $p^{2}\rightarrow\infty$ that corresponds to
the UV regime $p^{2}\gg M^{2}\gg m^{2}$. In this case,
Eqs.~ (\ref{rDG_gaze}), (\ref{rDG_ze2}), and  (\ref{rDG_ga2})
simplify and we arrive at the expressions

\beq
\Si^{\textrm{UV}}_{\ze^{2}}(p^{2}\rightarrow\infty)
&=&
\frac{i\ze^{2}p^{4}}{(4\pi)^{2}\theta^{4}}\,\Bigg\{5
\bigg[\frac{1}{\ep}-\ln\bigg(\frac{p^{2}}{\mu^{2}}\bigg)\bigg]
+ 3 - \frac{15(M^{2}+m^{2})}{p^{2}}
\nn
\\
&&
+\,\,\frac{10(m^{4}+m^{2}M^{2}+M^{4})}{p^{4}}
\ln\Big(\frac{p^{2}}{M^{2}}\Big)
+ \frac{35(M^{2}+m^{2})}{6p^{4}}
\nn
\\
&&
+\,\, \frac{40M^{2}m^{2}}{3p^{4}} +\frac{10m^{6}}{p^{4}M^{2}}\ln\Big(\frac{m^{2}}{M^{2}}\Big)
+ \mathcal{O}\Big(\frac{M^6}{p^6}\Big) \bigg\},
\eeq
\beq
\Si^{\textrm{UV}}_{\ga\ze}(p^{2}\rightarrow\infty)
&=&
-\, \frac{i\ga\ze p^{2}}{(4\pi)^{2}\theta^{4}}
\,\Bigg\{3\bigg[\frac{1}{\ep}-\ln\Big(\frac{p^{2}}{\mu^{2}}\Big)\bigg]
+ 7 - \frac{9(M^{2}+m^{2})}{p^{2}}
\nn
\\
&&
-\,\,
\frac{6(M^{2}+m^{2})}{p^{2}}\ln\Big(\frac{p^{2}}{M^{2}}\Big)
- \frac{6m^{4}}{p^{2}M^{2}}\ln\Big(\frac{m^{2}}{M^{2}}\Big)
+ \mathcal{O}\Big(\frac{M^4}{p^4}\Big) \bigg\},
\eeq
\beq
\Si^{\textrm{UV}}_{\ga^{2}}(p^{2}\rightarrow\infty)
&=& \frac{i\ga^{2}}{(4\pi)^{2}\theta^{4}}\,
\Bigg\{2\bigg[\frac{1}{\ep}
+ \ln\Big(\frac{p^{2}}{\mu^{2}}\Big)
+ 2\ln\Big(\frac{\mu^{2}}{M^{2}}\Big)\bigg] + 5
\nn
\\
&& +\,\,
\frac{4m^{2}}{M^{2}}\ln\Big(\frac{m^{2}}{M^{2}}\Big)
+ \mathcal{O}\Big(\frac{M^{2}}{p^{2}}\Big)\bigg\}.
\eeq
As expected in the UV regime,
the leading logarithmic terms in the
form factor, i.e., the terms with $\ln\big(p^{2}/\mu^{2}\big)$,
are proportional to the corresponding divergences. 
It is
easy to verify that, when returning to the coordinate representation,
the divergent part of the expressions above, together with
\eqref{1la} in the UV, correspond to the result \eqref{div_4},
obtained from the heat-kernel technique.

On the other hand, assuming
$m^{2} = 0$ in the formulas of $\Si_{\ze^{2}},\; \Si_{\ga\ze}$ and
$\Si_{\ga^{2}}$ (see part \ref{secB2} of Appendix \ref{secB}), the 
analysis of the IR regime $M^{2}\gg p^2$ of these corrections 
provides
\beq
\label{IRze2}
\Si^{\textrm{IR}}_{\ze^{2}}(M^{2}\gg p^{2})\bigg|_{m^{2}=0}
&=& \frac{i\ze^{2}p^{4}}{(4\pi)^{2}\th^{4}}\,
\Bigg\{5\bigg[\frac{1}{\ep}
+ \ln\Big(\frac{\mu^{2}}{M^{2}}\Big)\bigg]
- \frac{1}{6}\bigg(7+\frac{35p^{2}}{2M^{2}}
-\frac{9p^{4}}{2M^{4}}\bigg)
\nn
\\
&&
+ \,\, \frac{p^{4}}{2M^{4}}\ln\Big(\frac{M^{2}}{p^{2}}\Big)
+ \mathcal{O}\Big(\frac{p^{6}}{M^{6}}\Big) \bigg\},
\eeq
\beq
\label{IRgaze}
\Si^{\textrm{IR}}_{\ga\ze}(M^{2}\gg p^{2})\bigg|_{m^{2}=0}
&= &
-\frac{i\ga\ze p^{2}}{(4\pi)^{2}\th^{4}}\,
\Bigg\{3\bigg[\frac{1}{\ep}
+ \ln\Big(\frac{\mu^{2}}{M^{2}}\Big)\bigg]
- \frac12 +\frac{2p^{2}}{3M^{2}}
\nn
\\
&&
- \,\,\frac{p^{4}}{M^{4}}
\bigg[\frac{7}{20} - \frac12 \ln\Big(\frac{p^{2}}{M^{2}}\Big)\bigg]
+ \mathcal{O}\Big(\frac{p^6}{M^6}\Big)\bigg\},
\eeq
\beq
\label{IRga2}
\Si^{\textrm{IR}}_{\ga^{2}}(M^{2}\gg p^{2})\bigg|_{m^{2}=0}
& = & \frac{i\ga^{2}}{(4\pi)^{2}\th^{4}}
\,\Bigg\{2\bigg[\frac{1}{\ep}
+\ln\bigg(\frac{\mu^{2}}{M^{2}}\bigg)\bigg]
+\frac{13}{6}\frac{p^{2}}{M^{2}}
\nn
\\
&&
-\,\, \frac{p^{4}}{2M^{4}}\bigg[\frac{8}{5}
+\ln\Big(\frac{p^{2}}{M^{2}}\Big)\bigg]
+ \mathcal{O}\bigg(\frac{p^6}{M^6}\bigg)\bigg\}.
\eeq
The last formulas show that, in the IR limit, the divergences and
momentum dependence do not correlate with each other, exactly
as it is expected \cite{AC} (see also \cite{apco,fervi,Omar-FF4D}
for the semiclassical theory). We have found that this basic
feature holds also for the ``mixed'' diagrams, such that the
Appelquist-Carazzone theorem is valid for the fourth-derivative
model with non-polynomial interactions.
In the expressions \eqref{IRze2} and \eqref{IRgaze},
the nonlocal part with logarithmic form factor
$\ln\big(p^{2}/M^{2}\big)$ is suppressed by powers of $M^2$,
whereas in \eqref{IRga2} this is not the
case, as the factor $\ga^2$ (remember that $\ga=\th^2M^2$)
cancels this suppression in the terms proportional to $p^4$.
Although it may not be obvious, one can check that these nonlocal 
structures represent the contributions from the light sector alone.
The one-loop diagrams with mixed (light and massive ghost) internal
lines and (of course) the pure ghost contributions collapse and
produce only trivial dependencies on the external momentum.
All in all, we verified the quadratic decoupling of the heavy mode
in the Feynman diagrams with the mixed contents.

\section{One-loop corrections in the effective theory}
\label{sec5}

The last element of our investigation is the comparison between
what remains from the logarithmic form factors of the full theory
in the IR and the leading logarithms in the effective (initially
local) theory without heavy degrees of freedom. According to the
existing expectations \cite{Don-94}, the two expression  should
demonstrate a perfect correlation. This result would mean, in
particular, that the quantum general relativity can serve as a
universal low-energy model in any renormalizable or
superrenormalizable approach to quantum gravity.

So, let us evaluate the quantum corrections to the propagator
in the effective low-energy model of \eqref{eff_action2},
containing only the light mode. We consider a scenario in which
the energy scale is much smaller than the Planck mass. Therefore,
we can assume that the EH and cosmological constant terms
dominate over the higher derivative terms, leaving only the
part $\mathcal{L}_{\textrm{IR}}$. Under these considerations,
the tree-level propagator of the conformal factor boils down to
\beq
\label{prop_eff}
\widetilde{G}_{\textrm{eff}}(k)
\,=\,-\,\frac{i}{2\ga\big(k^{2}-m^{2}\big)},
\eeq
where $m^2$ is defined in \eqref{gpro}. The vertices are the
same as those in 
\eqref{3la} and \eqref{4la}. Since
we are dealing with an effective model, we are not concerned that
$\mathcal{L}_{\textrm{IR}}$ is non-renormalizable, as we may
ignore the higher-derivative divergences\footnote{According to
the logic of the pioneer work \cite{Don-94} (see also \cite{Burgess})
the divergences in quantum gravity are local expressions and,
therefore, have no direct relation to the long-distance regime
corresponding to the IR limit.}.
Thus, our interest is to explore the contributions to the
cosmological constant and the Einstein-Hilbert terms.
These formulas can be compared with the structures found in the
IR limit of the full theory \eqref{eff_action2}.

In the low-energy effective theory, the relevant contribution
is given by
\beq
\Si^{\textrm{eff}}_{\ga^{2}}(p)
& = & \frac{ip^{4}}{(4\pi)^{2}}\biggl\{
\Big(\frac12-\frac54\,a+\frac38\,a^2\Big)
\,\bigg[\frac{1}{\ep} - \ln\Big(\frac{\mu^{2}}{m^{2}}\Big)\bigg]
+ \Big(1-\frac74\,a+\frac12\,a^{2}\Big)
\nn
\\
&&
-\,\,
\frac{1}{2c}\Big(\frac14\,a^{2}
-\frac{1}{c^{2}}+2\Big)\ln\Big(\frac{1+c}{1-c}\Big)\biggr\}.
\label{Sigma-IR}
\eeq
To make the comparison more explicit, consider the particular
case $\La=0$ (or, equivalently, $p^2 \rightarrow \infty$).
Then the last expression reduces to a simpler form,
\beq
\label{gaeff}
\Si^{\textrm{eff}}_{\ga^{2}}(p)\Big|_{\La=0}
&=& \frac{ip^{4}}{2(4\pi)^{2}}
\bigg[\frac{1}{\ep} - \ln\Big(\frac{p^{2}}{\mu^{2}}\Big) + 2\bigg].
\eeq
Note that the nonlocal contribution
$\ln\big(p^{2}/M^{2}\big)$ involving the interaction vertex
$\ga^{2}$ in the IR regime of the ``fundamental'' theory
\eqref{eff_action2}, Eq.~\eqref{IRga2}, 
correlates with the logarithmic
term of the result \eqref{gaeff}, regardless in the IR of
the  fundamental theory there is no UV divergence. From
these results, it is possible to establish the one-loop match
between these two scenarios, i.e., fundamental and effective.
Let us note that the emergence of $p^4$
factor in the effective approach and its identification with the
cosmological constant term has been discussed in the recent
literature. In particular, this issue was described in detail,
exactly as a reaction to the attribution of the part of the
gravitational form factor to the cosmological constant in
\cite{don22}.  It reality, the corresponding terms appear as
part of the expansion of the nonlocal form factors of the
$R_{\mu\nu}R^{\mu\nu}$ and $R^2$ terms and have no direct
relation to the cosmological constant term \cite{2simpQG}.

In order to establish the match between the fundamental and 
effective scenarios, we introduce the relation
\beq
\label{match}
\Si^{\textrm{IR}}_{\ga^{2}}
&=& \Si^{\textrm{eff}}_{\ga^{2}}
+\de^{\textrm{eff}}_{\ga^{2}},
\eeq
where $ \de^{\textrm{eff}}_{\ga^{2}} $ is an additional term which
represents, at the one-loop level, the difference between the correction
of the fundamental theory in the low energies and the correction of
the effective theory. Considering the collapse of the diagrams with
massive ghost internal lines in the IR regime of the fundamental
theory, as we saw in the previous section, we can identify that the
additional term in \eqref{match} is composed of contributions arising
from the collapse of loops in the mixed sector. These collapsed
diagrams reduce to the tadpole-type graphs, and the remaining part,
related to pure ghost loops, i.e.,
\beq
\label{match_delta}
\de^{\textrm{eff}}_{\ga^{2}}
&=&
c^{(1)}_{\ga^{2},\,\textrm{mixed}}
+ c^{(1)}_{\ga^{2},\,\textrm{ghost}}.
\eeq
The contribution of the tadpole-type in \eqref{match_delta} is
proportional to the mass $ m^{2} $, and hence vanishes in the
simplification adopted for the IR, namely assuming
$c^{(1)}_{\ga^{2},\,\textrm{mixed}}\big|_{m^{2}=0}=0$.
Using the results \eqref{IRga2} and \eqref{gaeff} in relation 
\eqref{match}, we find
the leading logarithmic terms in the form
\beq
\label{match2}
\de^{\textrm{eff}}_{\ga^{2}}\Big|_{m^{2}=0}
& = & \frac{i}{(4\pi)^{2}}
\,\Bigg\{2M^{4}\ln\bigg(\frac{\mu^{2}}{M^{2}}\bigg)
+\frac{13}{6}p^{2}M^{2}
-\frac{p^{4}}{2}\bigg[\frac{18}{5}
+\ln\Big(\frac{\mu^{2}}{M^{2}}\Big)\bigg]\Bigg\}.
\eeq
In the above expression we present only the
finite part, since the divergences can be removed by a suitable
renormalization procedure. As it should be expected,
the nonlocal part with momentum-dependent logarithmic
form factor is canceled and the IR matching condition, which
ensure the equivalence with the result \eqref{IRga2}, is satisfied
with $\de^{\textrm{eff}}_{\ga^{2}}$ contains only terms with
trivial dependencies on the external momentum.

\section{Implications for the cosmological constant problem}
\label{secCC}

The cosmological constant problem is one of the main unsolved
issues in the present-day theoretical physics. The problem was
formulated by Weinberg in \cite{Weinberg89} as the need to
explain the extremely precise fine tuning between the original
cosmological constant density in the vacuum action and the
huge induced contributions. One can reformulate the problem in
terms of the renormalization of the vacuum term \cite{CC-nova}
but this does not help too much in its resolution. There are also
many other interesting aspects of the problem, related to cosmology
(see, e.g., \cite{Padm-03,SahniStar-2006}). Along with the main
problem, in the quantum field theory framework we need to
understand whether the cosmological constant density and the
Newton constant are really constants or these parameters can
be slowly varying with the energy scale, as predicted, e.g.,
by the four-derivative model of \cite{antmot} (see also the examples
of discussions based on the extended models
\cite{AntoMzMtt94,Eliz94_OneLoop,AntoOdin94} and supersymmetric
generalization \cite{SusyConf96})
and, of course, in the full fourth-derivative
\cite{frts82,avbar86,Gauss} or even
higher-derivative models \cite{SRQG-betas} of quantum gravity.

As we saw in the previous sections, the naive Minimal Subtraction
- based approach to the renormalization group for the cosmological
constant term is not operational, as it ignores the decoupling of the
massive (ghost or healthy, in some models) degrees of freedom.
Assuming that all massive degrees of freedom have typical masses
of the Planck order of magnitude, all the cosmological applications
occur at the deep IR, where the Appelquist-Carazzone - type
decoupling changes the beta functions. The question is what remains
from these beta functions in the theory with both massive and
massless degrees of freedom \cite{Gauss,Polemic}?

The result which we got for the quantum theory of conformal factor
is that, in the deep IR, there remain the contributions  (\ref{IRze2}),
(\ref{IRgaze}) and (\ref{IRga2}), which fit the ones of effective
low-energy quantum theory based on the local model.
This provides a positive answer to one of the main questions posed in
\cite{Polemic} and confirms the hypothesis of \cite{Don-94}. We got
a strong confirmation that the IR limit of a higher derivative model
of quantum gravity, which has massive degrees of freedom (those may
be healthy modes or ghosts) corresponds to the loop corrections
in the effective quantum gravity based on general relativity. The
result comes from the model of \cite{antmot} which has
non-polynomial interactions, very close to the case of real models
of quantum gravity. In this sense, the new confirmation is more
explicit than the results obtained previously in the framework of
simplified models, such as \cite{FF-19}).
Looking at the remnant expressions of the form
factors in the  IR, we note that the terms without $p^4$ or $p^2$
have only $\log (\mu^2/M^2)$ coefficients and no
momentum-dependent logarithmic terms. This is consistent with
the previous analysis of the possible quantum corrections, indicating
that the logarithmic (in Euclidean momentum) form factors cannot
be inserted in the cosmological constant term \cite{apco}. Recently,
similar observations were done, e.g., in
\cite{Mottola-2017,Sola22,Mottola22}). 
This situation confirms the general
expectation that there cannot be \textit{physical} running of the
cosmological constant term, detectable by means of flat-space
calculations, as discussed in \cite{apco} and more recently in
\cite{2simpQG}. Let us stress that this does not mean that the
cosmological constant running is impossible in general, it just
cannot be detected in the flat-space calculations \cite{DCCrun}.

It is worth noting that one can observe the running of the
cosmological constant term using the non-covariant parametrization
such as (\ref{metpar}), just as a decoupling in the beta function of
the $\phi^4$-interaction. Up to a certain extent, the corresponding
calculations were already developed in \cite{bexi} and can be
generalized to other theories, including quantum gravity. This would
be certainly an interesting way to extend the present work. However,
it is important to be careful with the expectations to the results of
such an extension, as there will always remain a question about the
physical interpretation of the result obtained by means of
non-covariant methods.

\section{Conclusions and discussions}
\label{sec6}

We have considered the detailed renormalization in the theory of
the conformal factor. Assuming a small cosmological constant, such
a theory possesses two mass scales with a strong hierarchy between
them. On top of that, the theory has non-polynomial interactions
and is renormalizable \cite{antmot}. These features make the model
qualitatively similar to the higher derivative quantum gravity.
Previously, the one-loop calculations in this model were performed
in the Minimal Subtraction scheme and we performed a more detailed
analysis in the momentum subtraction scheme of renormalization.

The analysis of the nonlocal form factors shows that in the UV, we
meet a correspondence with the Minimal Subtraction scheme results.
On the other hand, in the IR we met a strong deviation from this
simplified scheme and, as it was anticipated, a good agreement
with the calculation in the effective model that ignores the massive
degree of freedom. One of the new details is that the ``mixed''
diagrams, with the internal lines of both small-mass and large-mass
fields, transform into tadpoles. These diagrams contribute to the
UV divergences, but not to the nonlocal form factors.
This means, the diagrams with the large-mass internal lines
collapse and become irrelevant in the IR. In particular, nonlocal
structures that survive in this regime and contribute to the
cosmological constant sector, are correlated with the UV divergent
part in an effective version of the model containing only the light
mode.

It would be certainly interesting to extend the analysis which was
presented above, for the models of ``real" quantum gravity, i.e., the
theory of quantum metric. As we mentioned in the Introduction,
this is a technically more challenging problem because such a
theory has gauge invariance and complicated tensor structures in
the sectors of quantum metric and ghost. However, the results
presented above show that there are very good chances to meet
the expectation of  universality of quantum general relativity as
an effective theory of quantum gravity, at least in the fourth
derivative \cite{Stelle77} and, probably, all polynomial models
introduced in \cite{highderi}, where all extra degrees of freedom
have the masses of the Planck order of magnitude \cite{ABSh2}.

At the same time, the situation may be more complicated in the
non-local theories of quantum gravity
\cite{Krasni_87,Kuz_89,Tombou_97,Modesto-nonloc}
(many further references can be found in the last review).
The most popular version of nonlocal models are free from
massive degrees of freedom at the tree-level. On the other hand,
starting from the one-loop level, the structure of the propagator
changes and there are infinitely many complex-energy and
complex-mass ghost-like states with the quasi-continuous
mass spectrum \cite{CountGhosts}. In this case, the universality
of general relativity as the IR quantum gravity theory is rather
uncertain. This means, there are still many interesting issues to
explore in the area of the present work.

The last point is that we have found a good correspondence between
the IR limit of the theory with massless and large-mass degrees of
freedom and the UV limit of the effective theory without the massive
particles. This correspondence extends to the contributions in the
cosmological constant sector of the gravitational action. However,
these contributions are not momentum-dependent, confirming the
general no-go statement \cite{apco} concerning the detection
of the cosmological constant running by means of the flat-space
calculations. On the other hand, this output does not means that
such a running is impossible by itself. Instead, it should be
interpreted as a challenge to develop new methods of effective
field theory calculations which would be appropriate for
clarifying this issue.

\section*{Acknowledgments}

W.C.S. is grateful to CAPES for supporting his PhD project.
The work of I.Sh. is partially supported by Conselho Nacional de
Desenvolvimento Cient\'{i}fico e Tecnol\'{o}gico - CNPq under the
grant 303635/2018-5.

\newpage

\appendix
\section*{Appendices}
\addcontentsline{toc}{section}{Appendices}
\renewcommand{\thesubsection}{\Alph{subsection}}

\subsection{Feynman diagrams}
\label{secA}

In this appendix, we present the set of one-loop
Feynman diagrams that correspond to the corrections
to the two-point function.
\begin{figure}[H]
\centering
\includegraphics[scale=0.54]{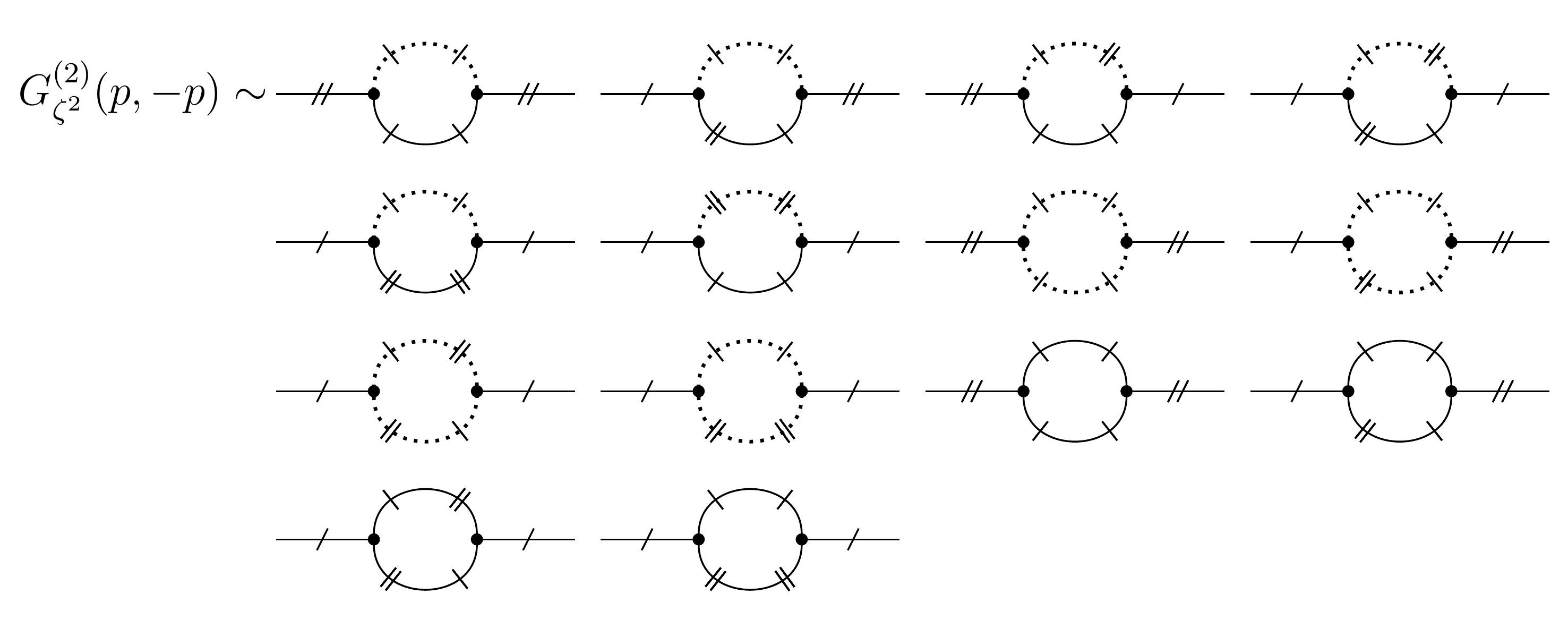}
\vspace{-9mm}
\caption{Diagrams for the two-point function that
provide one-loop contributions to the renormalization
of the coupling $ \ze $.
Solid lines indicate light degrees of freedom,
dashed lines stand for the massive ghosts, and
the primes denotes derivatives acting on the
propagators.} \label{fig2}
\end{figure}
\begin{figure}[H]
\centering
\includegraphics[scale=0.37]{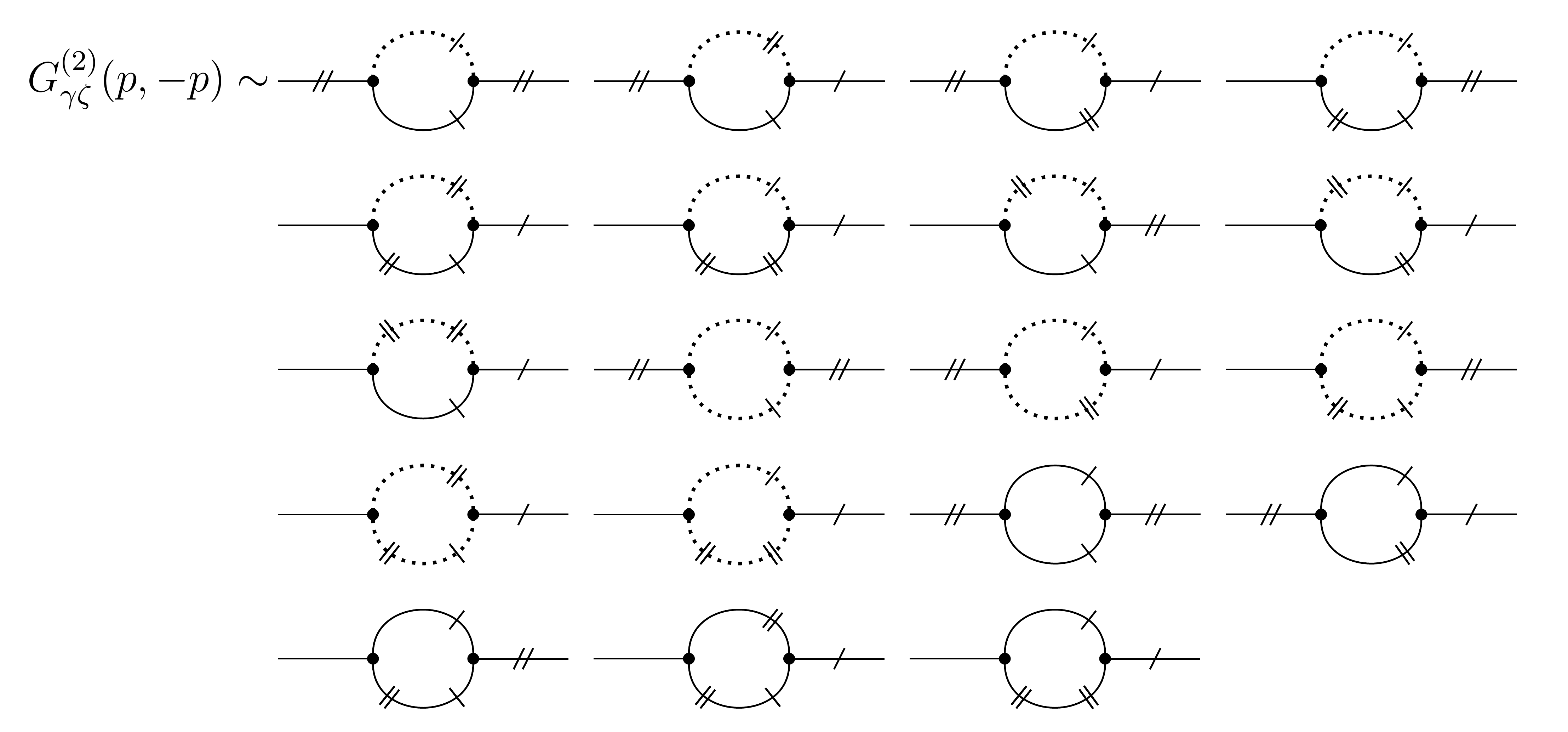}
\vspace{-8mm}
\caption{Diagrams for the two-point function with the interaction
vertex $ \ga\ze $.}
\label{fig3}
\end{figure}
\begin{figure}[H]
\centering
\includegraphics[scale=0.54]{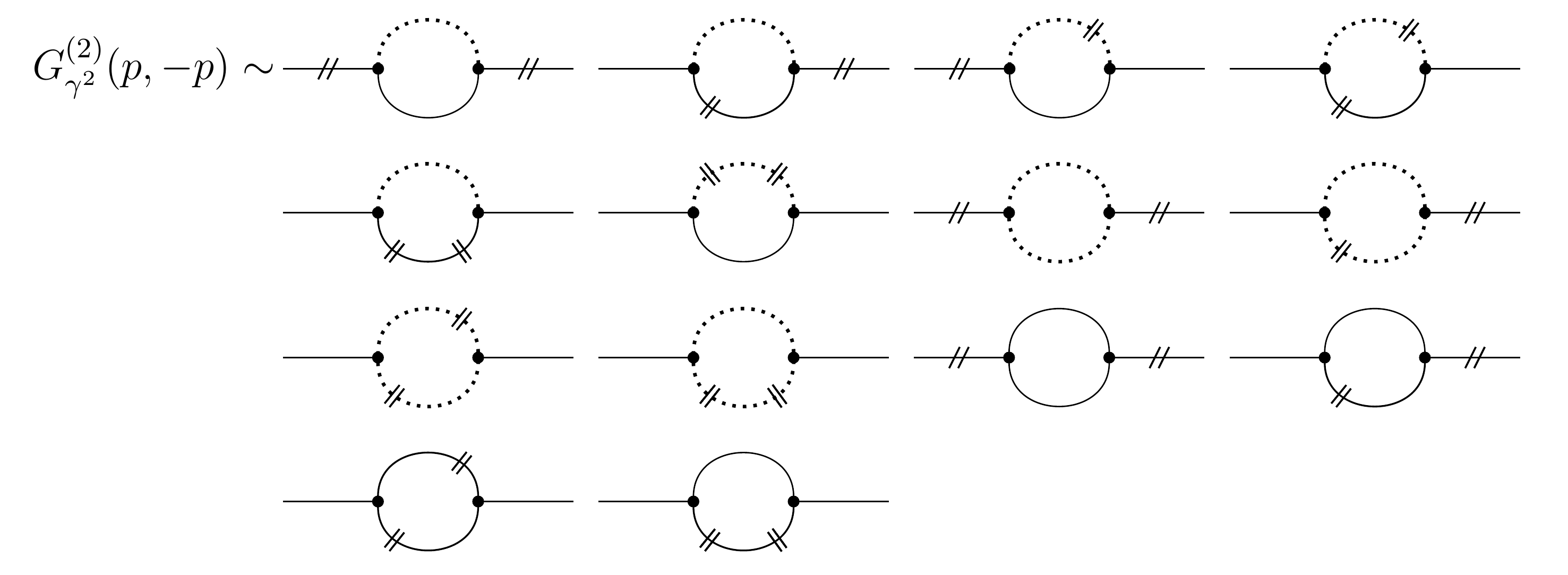}
\vspace{-8mm}
\caption{Diagrams for the two-point function with
the interaction vertex $ \ga^{2} $.}
\label{fig4}
\end{figure}
\begin{figure}[H]
\centering
\includegraphics[scale=0.165]{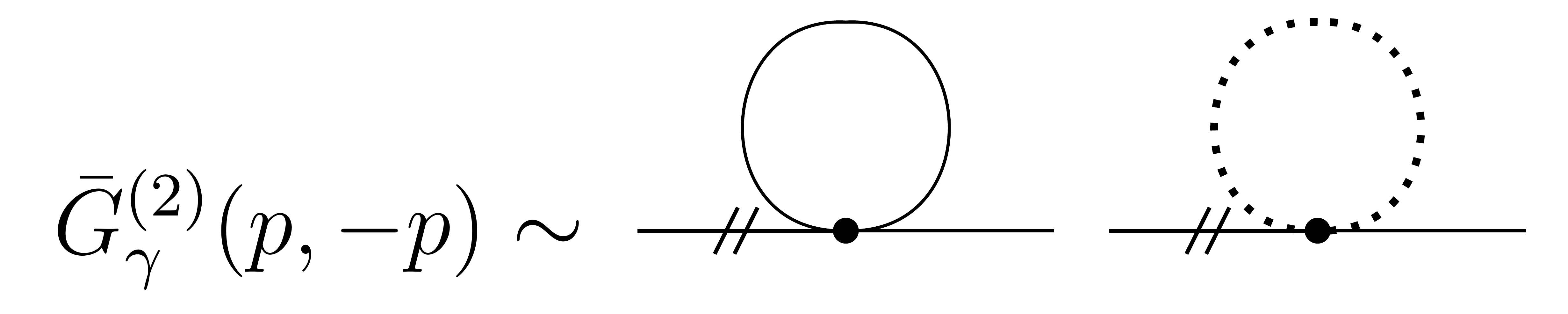}
\vspace{-3mm}
\caption{Diagrams for the two-point function with the interaction
vertex $ \ga $.}
\label{fig5}
\end{figure}
\begin{figure}[H]
\centering
\includegraphics[scale=0.16]{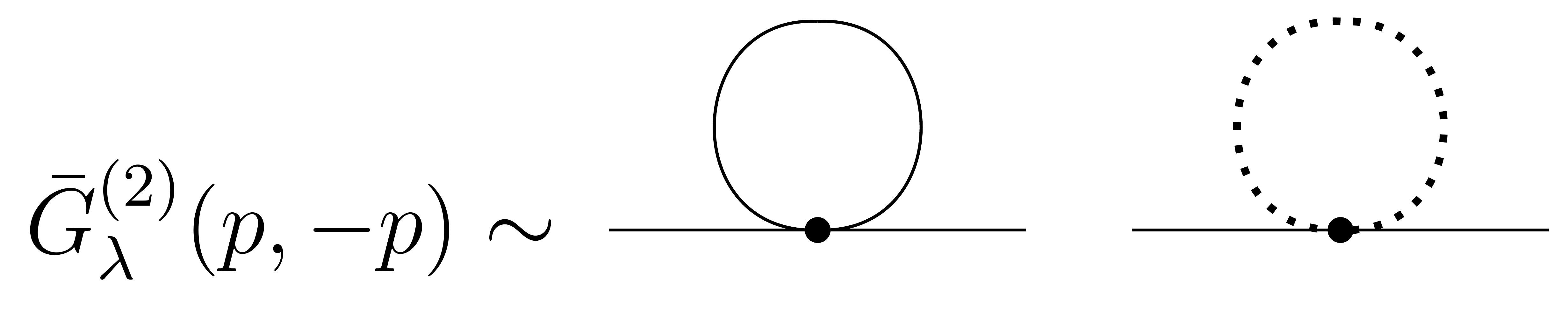}
\vspace{-3mm}
\caption{Diagrams for the two-point function with the
interaction vertex $ \la $.} \label{fig6}
\end{figure}
\begin{figure}[H]
\centering
\includegraphics[scale=0.27]{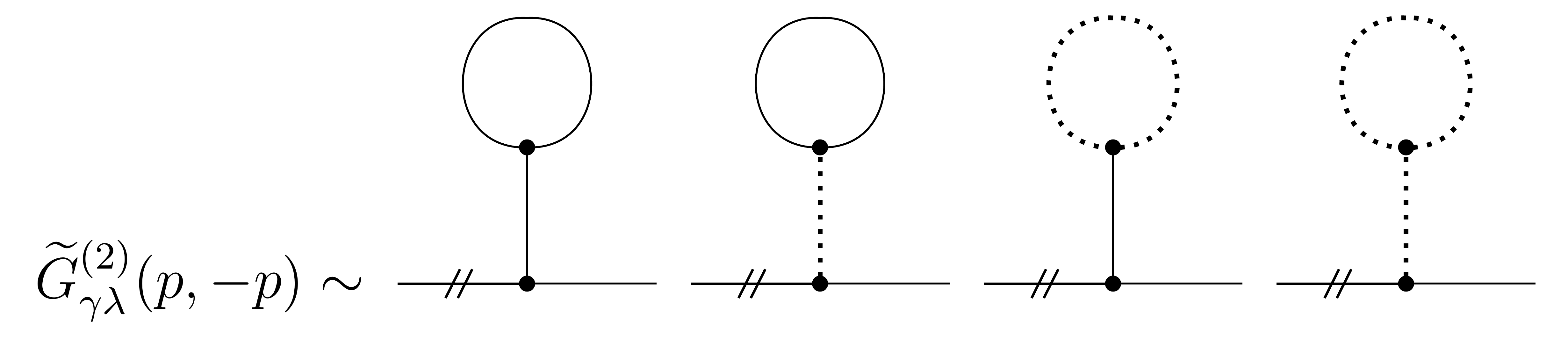}
\vspace{-2mm}
\caption{Diagrams (tadpole) for the two-point function
with the interaction vertex $ \ga\la $.}
\label{fig7}
\end{figure}
\begin{figure}[H]
\centering
\includegraphics[scale=0.27]{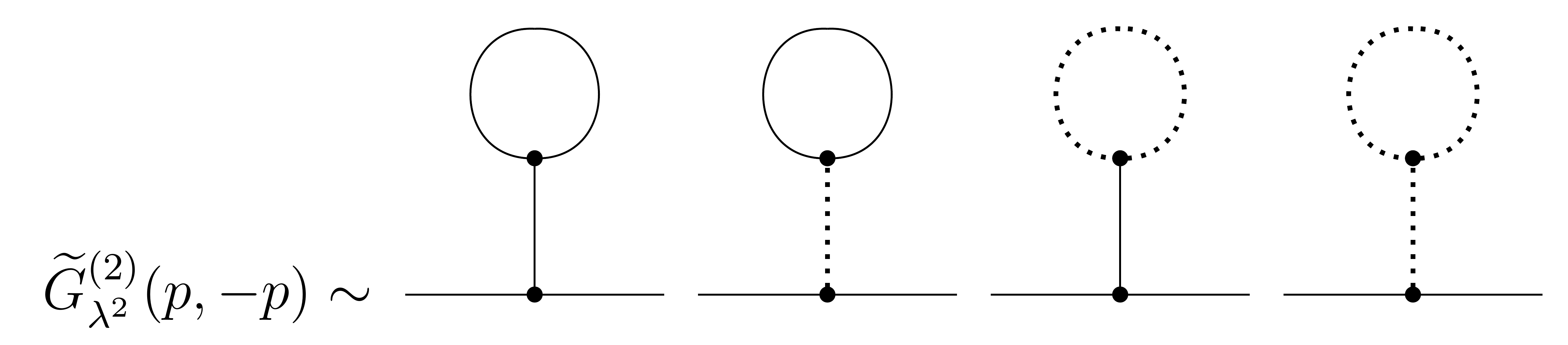}
\vspace{-2mm}
\caption{Diagrams (tadpole) for the two-point function
with the interaction vertex $ \la^{2} $.}
\label{fig8}
\end{figure}

\begin{figure}[H]
\centering
\includegraphics[scale=0.368]{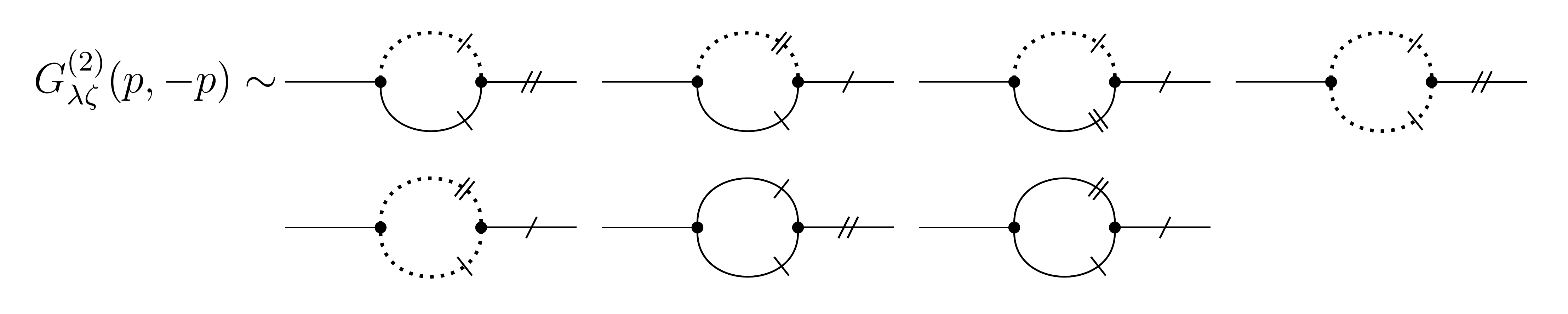}
\vspace{-9mm}
\caption{Divergence-free diagrams for the two-point
function with the vertex $ \la\ze $.}
\label{fig15}
\end{figure}

\begin{figure}[H]
\centering
\includegraphics[scale=0.368]{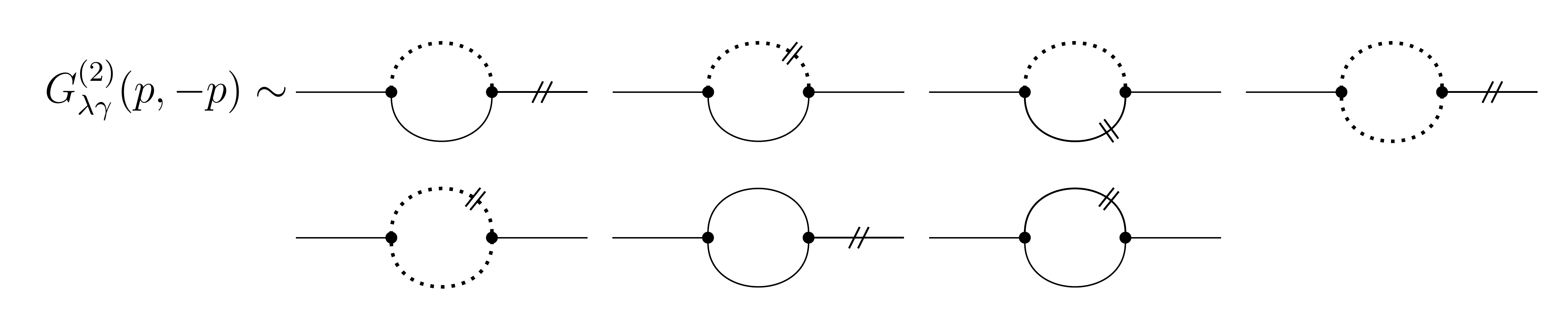}
\vspace{-8mm}
\caption{Divergence-free diagrams for the two-point
function with the vertex $ \la\ga $.}
\label{fig16}
\end{figure}

\begin{figure}[H]
\centering
\includegraphics[scale=0.195]{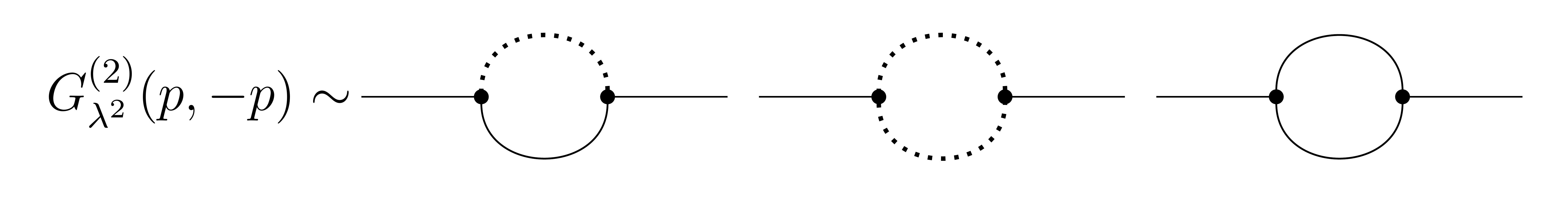}
\vspace{-4mm}
\caption{Divergence-free diagrams for the two-point
function with the vertex $ \la^{2} $.}
\label{fig17}
\end{figure}

Diagrams that contribute only finite corrections to the
two-point function are shown in Figs.~\ref{fig15},
\ref{fig16} and \ref{fig17}.

\subsection{Intermediate results}
\label{secB}

In this appendix, we present some intermediate results related
to the calculation of Feynman integrals in sect. \ref{sec3}.

\subsubsection{Feynman integrals}
\label{secB1}

By using the Feynman parametrization
\beq
\frac{1}{ab} \,=\, \int_{0}^{1}\frac{dx}{\big[(a-b)x+b\big]^{2}},
\eeq
and performing the following shift of integration variable
$ k=q+px $, we can rewrite the integrals related to the mixed sector
in the expressions \eqref{DG_ze2}, \eqref{DG_gaze} and \eqref{DG_ga2},
respectively, as
\beq
\Si^{(2\om)}_{\textrm{mixed},\,\ze^{2}}(p) & = & -\,\frac{2\ze^{2}\,p^{4}}{\theta^{4}(m^{2}-M^{2})^{2}}
\frac{(4\om^{2}-1)}{\om(1+\om)}\int^{1}_{0} dx\,I_4,
\eeq
\beq
\Si^{(2\om)}_{\textrm{mixed},\,\ga\ze}(p) & = & -\,\frac{2\ga\ze\,p^{2}}{\theta^{4}(m^{2}-M^{2})^{2}}
\frac{(2\om-1)}{\om}\int^{1}_{0} dx\,\Big[
\big(x^{2}-x+1\big)p^{2}I_2+I_4\Big]
\eeq
and
\beq
\Si^{(2\om)}_{\textrm{mixed},\,\ga^{2}}(p) & = &  -\,\frac{2\ga^{2}}{\theta^{4}(m^{2}-M^{2})^{2}}
\int^{1}_{0} dx\,\bigg[I_4
+\frac{1+4\om+4\big(x^{2}-x\big)(\om+1)}{2\om}p^{2}I_2
\nn
\\
&& +\,\big(x^{2}-x+1\big)^{2}p^{4}I_1\bigg],
\eeq
where, in Minkowski space,
\beq
&& I_{1}
\,=\,
\int \frac{d^{2\om}q}{(2\pi)^{2\om}}
\frac{1}{(q^{2}-\De)^{2}}
\,=\,
\frac{i}{(4\pi)^{\om}}\Ga(2-\om)\De^{\om-2},
\\
&& I_{2}
\,=\,
\int \frac{d^{2\om}q}{(2\pi)^{2\om}}\frac{q^{2}}{(q^{2}-\De)^{2}}
\,=\,
-\frac{i}{(4\pi)^{\om}}\om\Ga(1-\om)\De^{\om-1},
\\
&& I_{4}
\,=\,
\int \frac{d^{2\om}q}{(2\pi)^{2\om}}\frac{q^{4}}{(q^{2}-\De)^{2}}
\,=\,
\frac{i}{(4\pi)^{\om}}\om(1+\om)\Ga(-\om)\De^{\om}.
\eeq
Here we define $\De\equiv p^{2}x(x-1)+(M^{2}-m^{2})x+m^{2}$. For the integrals in the other sectors, we have the same results as above with $\De=\De_{\textrm{ghost}}=p^{2}x(x-1)+M^{2}$ in the case of the ghost sector, and $\De=\De_{\textrm{light}}=p^{2}x(x-1)+m^{2}$ for the light sector.

\subsubsection{
Corrections $\Si_{\ze^{2}},\; \Si_{\ga\ze}$ and $\Si_{\ga^{2}}$ 
for the case of $m^{2} = 0$}
\label{secB2}

The contributions from the diagrams shown in Figs.~\ref{fig2}, \ref{fig3} and \ref{fig4}, assuming
$m^{2} = 0$, can be written, respectively, as
\beq
\label{m0ze2}
\Si_{\ze^{2}}(p)\bigg|_{m^{2}=0}
&=& \frac{i\ze^{2}}{(4\pi)^{2}\th^{4}}
\,\Bigg\{\frac{5p^{4}}{\ep}
+\al^{(2)}_{\ze}(p)
\ln\Big(\frac{\mu^{2}}{M^{2}}\Big)
+\xi^{(2)}_{\ze}(p)
+\be^{(2)}_{\ze,\,\textrm{light}}(p)
\ln\Big(\frac{\mu^{2}}{p^{2}}\Big)
\nn
\\
&&
+ \,\, \be^{(2)}_{\ze,\,\textrm{ghost}}(p)
\ln \Big(\frac{1+d}{1-d}\Big)
+\be^{(2)}_{\ze,\,\textrm{mixed}}(p)
\ln\Big(\frac{M^{2}}{M^{2}+p^{2}}\Big)
\bigg\},
\eeq
\beq
\label{m0gaze}
\Si_{\ga\ze}(p)\bigg|_{m^{2}=0}
& = &
- \,\frac{i\ga\ze}{(4\pi)^{2}\th^{4}}
\,\Bigg\{\frac{3p^{2}}{\ep}
+\al^{(2)}_{\ga\ze}(p)
\ln\Big(\frac{\mu^{2}}{M^{2}}\Big)
+\xi^{(2)}_{\ga\ze}(p)
+\be^{(2)}_{\ga\ze,\,\textrm{light}}(p)
\ln\Big(\frac{\mu^{2}}{p^{2}}\Big)
\nn
\\
&&
+ \,\, \be^{(2)}_{\ga\ze,\,\textrm{ghost}}(p)
\ln \Big(\frac{1+d}{1-d}\Big)
+\be^{(2)}_{\ga\ze,\,\textrm{mixed}}(p)
\ln\Big(\frac{M^{2}}{M^{2}+p^{2}}\Big)
\bigg\}
\eeq
and
\beq
\label{m0ga2}
\Si_{\ga^{2}}(p)\bigg|_{m^{2}=0}
& = & \frac{i\ga^{2}}{(4\pi)^{2}\th^{4}}
\,\Bigg\{\frac{2}{\ep}
+\al^{(2)}_{\ga}(p)
\ln\Big(\frac{\mu^{2}}{M^{2}}\Big)
+\xi^{(2)}_{\ga}
+\be^{(2)}_{\ga,\,\textrm{light}}(p)
\ln\Big(\frac{\mu^{2}}{p^{2}}\Big)
\nn
\\
&&
+ \,\, \be^{(2)}_{\ga,\,\textrm{ghost}}(p)
\ln \Big(\frac{1+d}{1-d}\Big)
+\be^{(2)}_{\ga,\,\textrm{mixed}}(p)
\ln\Big(\frac{M^{2}}{M^{2}+p^{2}}\Big)
\bigg\},
\eeq
where $\be$'s are coefficients of the nonlocal part with momentum-dependent logarithmic form factor, decomposed according to the light, ghost and mixed sectors, while $\al$'s and $\xi$'s are coefficients of contributions involving the combination of different sectors:
\beq
\label{coefs}
&&
\be^{(2)}_{\ze,\,\textrm{light}}(p)
\,=\,
\frac{p^{8}}{2M^{4}},
\hspace{10mm}
\be^{(2)}_{\ga\ze,\,\textrm{light}}(p)
\,=\,
- \,\frac{p^{6}}{2M^{4}},
\hspace{10mm}
\be^{(2)}_{\ga,\,\textrm{light}}(p)
\,=\,
\frac{p^{4}}{2M^{4}},
\nn
\\
&&
\be^{(2)}_{\ze,\,\textrm{ghost}}(p)
\,=\,
- \,\frac{p^{8}}{2M^{4}d^{5}},
\hspace{36.7mm}
\be^{(2)}_{\ga\ze,\,\textrm{ghost}}(p)
\,=\,
- \,\frac{p^{2}(8M^{4}-2M^{2}p^{2}-p^{4})}{2M^{4}d},
\nn
\\
&&
\be^{(2)}_{\ga,\,\textrm{ghost}}(p)
\,=\,
- \,\frac{(2M^{2}-p^{2})^{2}}{2M^{4}d},
\hspace{26.5mm}
\be^{(2)}_{\ze,\,\textrm{mixed}}(p)
\,=\,
- \,\frac{(M^{2}+p^{2})^{5}}{M^{4}p^{2}},
\nn
\\
&&
\be^{(2)}_{\ga\ze,\,\textrm{mixed}}(p)
\,=\,
-\,\frac{(M^{2}-p^{2})(M^{2}+p^{2})^{3}}{M^{4}p^{2}},
\hspace{8mm}
\be^{(2)}_{\ga,\,\textrm{mixed}}(p)
\,=\,
-\,\frac{(M^{2}+p^{2})(M^{2}-p^{2})^{2}}{M^{4}p^{2}},
\nn
\\
&&
\al^{(2)}_{\ze}(p)
\,=\,
p^{4}\bigg(5-\frac{p^{4}}{2M^{4}}\bigg),
\hspace{16.1mm}
\al^{(2)}_{\ga\ze}(p)
\,=\,
p^{2}\bigg(3+\frac{p^{4}}{2M^{4}}\bigg),
\hspace{8mm}
\al^{(2)}_{\ga}(p)
\,=\,
2-\frac{p^{4}}{2M^{4}},
\nn
\\
&&
\xi^{(2)}_{\ze}(p)
\,=\,
7p^{4}
-\frac{9M^{2}p^{2}}{2}
-M^{4},
\hspace{8mm}
\xi^{(2)}_{\ga\ze}(p)
\,=\,
6p^{2}-M^{2},
\hspace{17.8mm}
\xi^{(2)}_{\ga}
\,=\,
3.
\eeq

\subsection{Results of the finite contributions 
$\Si_{\la\ze}$, $\Si_{\la\ga}$ and $\Si_{\la^{2}}$ }
\label{secC}

We collect here the results of the self-energy corrections that are free of divergences. The contributions from the diagrams shown in Figs.~\ref{fig15}, \ref{fig16} and \ref{fig17} are, respectively,
\beq
\label{rDG_laze}
\Si_{\la\ze}(p)
& = &
-\frac{16i\la\ze}{(4\pi)^{2}\th^{4}}\,
\bigg\{2
+\frac{1}{(ab)^{2}c^{3}}\ln\Big(\frac{1+c}{1-c}\Big)
+\frac{1}{(ab)^{2}d^{3}}\ln\Big(\frac{1+d}{1-d}\Big)
\nn
\\
&&
+ \,
\dfrac{1}{ab}\bigg(A^{2}+ab+\dfrac{a}{2}+2\bigg) \ln(1+4b)
-\frac{A^{3}}{(ab)^{2}}
\ln \Big[\frac{(A+1)^{2}-(ab)^{2}}{(A-1)^{2}-(ab)^{2}}\Big]\bigg\},
\eeq
\beq
\label{rDG_laga}
\Si_{\la\ga}(p)
& = &
\frac{16i\la\ga}{(4\pi)^{2}\th^{4}p^{2}}\,
\bigg\{
\frac{1}{(ab)^{2}c}\bigg(\dfrac{a}{2}-1\bigg)
\ln\Big(\frac{1+c}{1-c}\Big)
+\frac{1}{(ab)^{2}d}\bigg(2ab+\dfrac{a}{2}-1\bigg)
\ln\Big(\frac{1+d}{1-d}\Big)
\nn
\\
&&
+ \,
\bigg(1+\dfrac{1}{2b}\bigg) \ln(1+4b)
-\frac{A}{(ab)^{2}}\bigg(ab+\dfrac{a}{2}-1\bigg)
\ln \Big[\frac{(A+1)^{2}-(ab)^{2}}{(A-1)^{2}-(ab)^{2}}\Big]\bigg\}
\eeq
and
\beq
\label{rDG_la2}
\Si_{\la^{2}}(p)
& = &
-\frac{512i\la^{2}}{(4\pi)^{2}\th^{4}p^{4}}\,
\bigg\{
\dfrac{1}{ab}\ln(1+4b)
+\frac{1}{(ab)^{2}c}\ln\Big(\frac{1+c}{1-c}\Big)
+\frac{1}{(ab)^{2}d}\ln\Big(\frac{1+d}{1-d}\Big)
\nn
\\
&&
- \,
\frac{A}{(ab)^{2}}
\ln \Big[\frac{(A+1)^{2}-(ab)^{2}}{(A-1)^{2}-(ab)^{2}}\Big]\bigg\}.
\eeq

\subsection{One-loop corrections to the three- and four-point vertices}
\label{secD}

\begin{figure}[H]
\centering
\includegraphics[scale=0.295]{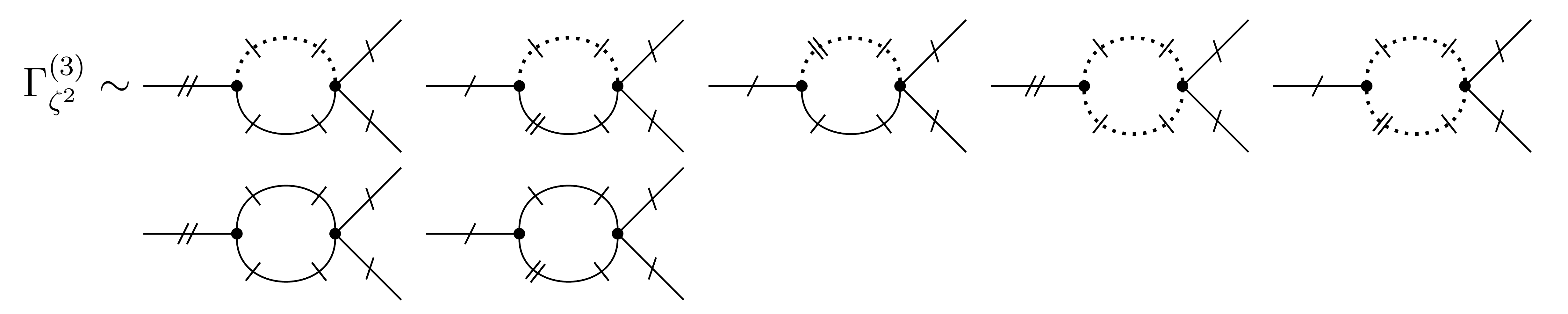}
\vspace{-3mm}
\caption{Diagrams for the three-point function with
the interaction vertex $\ze^{2}$.}
\label{fig9}
\end{figure}

\begin{figure}[H]
\centering
\includegraphics[scale=0.29]{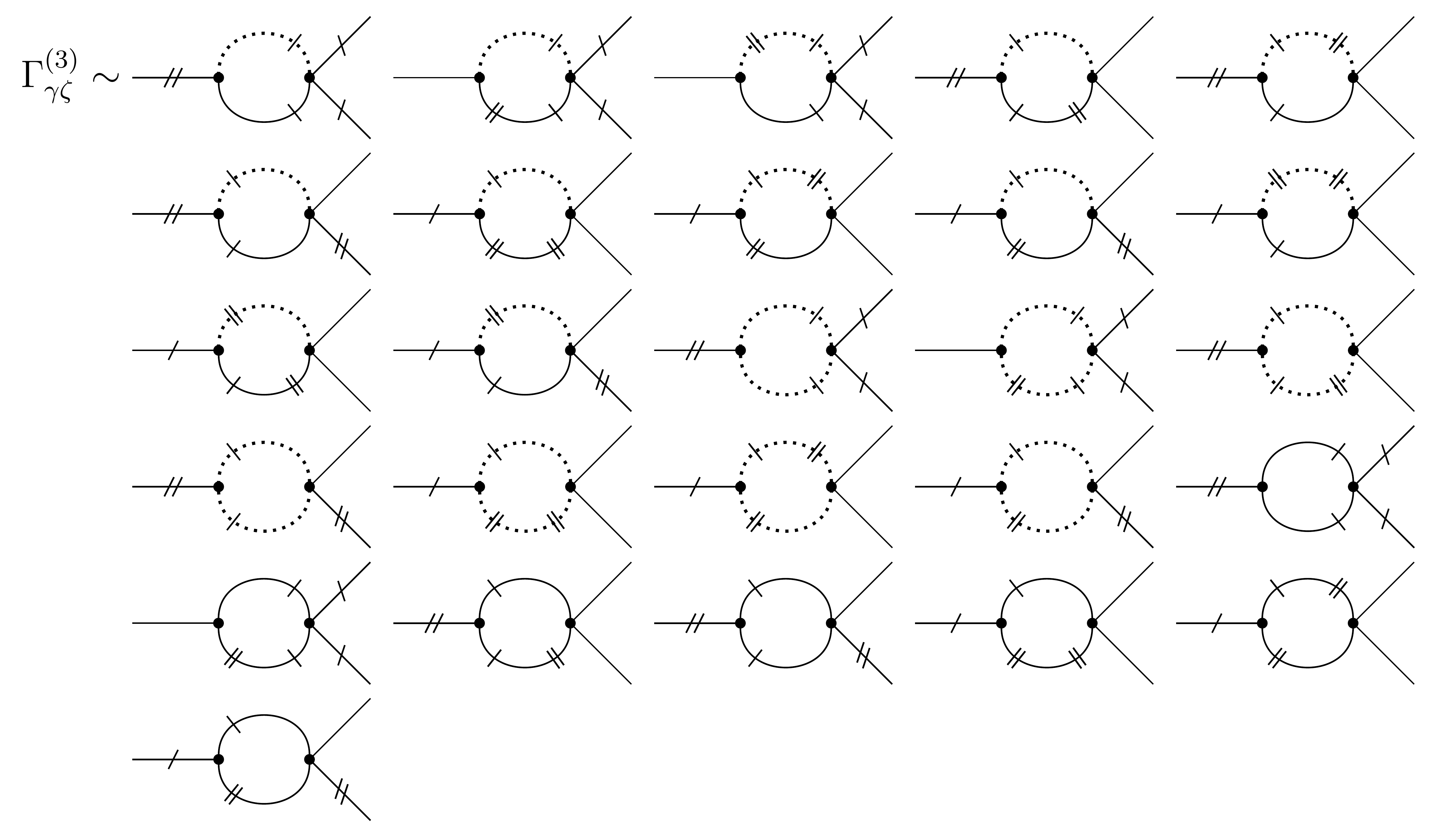}
\vspace{-3mm}
\caption{Diagrams for the three-point function with
the interaction vertex $\ga\ze$.}
\label{fig10}
\end{figure}

\begin{figure}[H]
\centering
\includegraphics[scale=0.298]{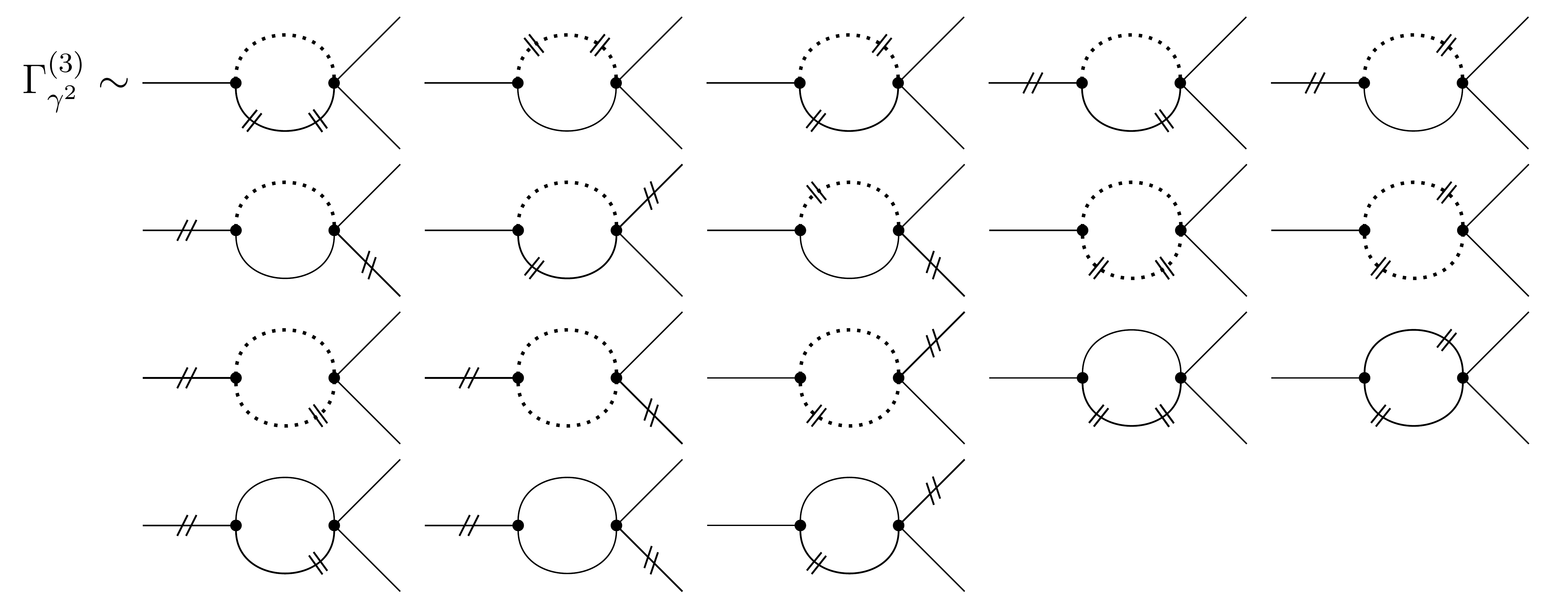}
\vspace{-3mm}
\caption{Diagrams for the three-point function with
the interaction vertex $\ga^{2}$.}
\label{fig11}
\end{figure}

This appendix is devoted to the one-loop corrections to the vertices.
In case of the three-point function, the relevant corrections are
associated with the diagrams in Figures \ref{fig9}, \ref{fig10}
and \ref{fig11}.
In the dimensional regularization scenario, these
corrections are given by following integrals,
\beq
\label{Vert34_ze2}
\Ga^{(3)}_{\ze^{2}}(p,r)\big|^{(2\om)}
& = &
-\,\frac{4\ze^{2}}{\th^{4}(m^{2}-M^{2})^{2}}
\int \frac{d^{2\om}k}{(2\pi)^{2\om}}\,\Ga^{(3,4)}_{\ze^{2}}
\biggl\{\frac{2}{(k^{2}-m^{2})\big[(k-p)^{2}-M^{2}\big]}
\nn
\\
&&
-\,\,\frac{1}{(k^{2}-m^{2})\big[(k-p)^{2}-m^{2}\big]}
\,-\,\frac{1}{(k^{2}-M^{2})\big[(k-p)^{2}-M^{2}\big]}
\biggr\}
\nn
\\
&&
+\,\, t\textrm{- and }u\textrm{-channel contributions},
\eeq
\beq
\label{Vert34_gaze}
\Ga^{(3)}_{\ga\ze}(p,r)\big|^{(2\om)}
& = &
\frac{2\ga\ze}{\th^{4}(m^{2}-M^{2})^{2}}
\int \frac{d^{2\om}k}{(2\pi)^{2\om}}
\,\Big[\Ga^{(3,4)}_{\ga\ze}+\Ga^{(3,4)}_{\ze\ga}\Big]
\biggl\{\frac{2}{(k^{2}-m^{2})\big[(k-p)^{2}-M^{2}\big]}
\nn
\\
&& -\,\,\frac{1}{(k^{2}-m^{2})\big[(k-p)^{2}-m^{2}\big]}
\,-\,\frac{1}{(k^{2}-M^{2})\big[(k-p)^{2}-M^{2}\big]}\biggr\}
\nn
\\
&& +\,\, t\textrm{- and }u\textrm{-channel contributions}
\eeq
and
\beq
\label{Vert34_ga2}
\Ga^{(3)}_{\ga^{2}}(p,r)\big|^{(2\om)}
& = &
-\,\frac{\ga^{2}}{\th^{4}(m^{2}-M^{2})^{2}}\int \frac{d^{2\om}k}{(2\pi)^{2\om}}\,\Ga^{(3,4)}_{\ga^{2}}
\biggl\{\frac{2}{(k^{2}-m^{2})\big[(k-p)^{2}-M^{2}\big]}
\nn
\\
&& -\,\,\frac{1}{(k^{2}-m^{2})\big[(k-p)^{2}-m^{2}\big]}
\,-\,\frac{1}{(k^{2}-M^{2})\big[(k-p)^{2}-M^{2}\big]}\biggr\}
\nn
\\
&& +\,\, t\textrm{- and }u\textrm{-channel contributions},
\eeq
where the combinations of the vertex factors
(for the $s$-channel diagrams) are
\beq
\label{vf34}
&&\Ga^{(3,4)}_{\ze^{2}} \,=\,
-2\big[k^{2}(p\cdot r)-k^{2}r^{2}
-2(k\cdot p-k\cdot r)(p\cdot r-k\cdot r)
-p^{2}(k\cdot r)+r^{2}(k\cdot p)\big]
\nn
\\
&&\hspace{1.45cm}
\,\,\times \big[(p\cdot k)^{2}-p^{2}k^{2}\big],
\nn
\\
&&\Ga^{(3,4)}_{\ga\ze}+\Ga^{(3,4)}_{\ze\ga} \,=\,
4\big[(k\cdot p)^{2}-k^{2}p^{2}\big]
\big(k^{2}-k\cdot p+p^{2}-p\cdot r+r^{2}\big)
-2\big[k^{2}(p\cdot r-r^{2})
\nn
\\
&&\hspace{1.45cm}
\,\,-p^{2}(k\cdot r)
-2(k\cdot r)(k\cdot r-p\cdot r)
+r^{2}(k\cdot p)
+2(k\cdot p)(k\cdot r-p\cdot r)\big]
\nn
\\
&&\hspace{1.45cm}
\,\,\times \big(k^{2}-k\cdot p+p^{2}\big),
\nn
\\
&&\Ga^{(3,4)}_{\ga^{2}} \,=\,
4\big[k^{2}-(k\cdot p)+p^{2}\big]
\big(k^{2}-k\cdot p+p^{2}-p\cdot r+r^{2}\big).
\eeq
Taking these integrals, we write the contributions to
the three-point function as
\beq
\label{vert34_ze}
\Ga^{(3)}_{\ze^{2}}(p,r)
& = & \frac{i\ze^{2}}{(4\pi)^{2}\theta^{4}}\,
\bigg\{
20\big[(p\cdot r)^{2}-p^{2}r^{2}\big]\bigg[\frac{1}{\ep}
+\ln\Big(\frac{\mu^{2}}{m^{2}}\Big)\bigg]
+\al^{(3)}_{\ze^{2}}(p,r)\ln(1+4b)
\nn
\\
&&
+\, \xi^{(3)}_{\ze^{2}}(p,r)
+\bigg[\be^{(3)}_{\ze^{2},\,\textrm{light}}(p,r)
\ln\Big(\frac{1+c}{1-c}\Big)
+\be^{(3)}_{\ze^{2},\,\textrm{ghost}}(p,r)
\ln \Big(\frac{1+d}{1-d}\Big)
\nn
\\
&&
+\, \be^{(3)}_{\ze^{2},\,\textrm{mixed}}(p,r)
\ln \Big[\frac{(A+1)^{2}-(ab)^{2}}{(A-1)^{2}-(ab)^{2}}\Big]
+(p\leftrightarrow -r)
+(p\leftrightarrow r-p)\bigg]
\bigg\}, \hspace{.8cm}
\eeq
\beq
\label{vert34_gaze}
\Ga^{(3)}_{\ga\ze}(p,r)
& = & -\frac{i\ga\ze}{(4\pi)^{2}\theta^{4}}\,
\bigg\{
6\big[p^{2}+r^{2}-(p\cdot r)\big]\bigg[\frac{1}{\ep}
+\ln\Big(\frac{\mu^{2}}{m^{2}}\Big)\bigg]
+\al^{(3)}_{\ga\ze}(p,r)\ln(1+4b)
\nn
\\
&&
+\, \xi^{(3)}_{\ga\ze}(p,r)
+\bigg[\be^{(3)}_{\ga\ze,\,\textrm{light}}(p,r)
\ln\Big(\frac{1+c}{1-c}\Big)
+\be^{(3)}_{\ga\ze,\,\textrm{ghost}}(p,r)
\ln \Big(\frac{1+d}{1-d}\Big)
\nn
\\
&&
+\, \be^{(3)}_{\ga\ze,\,\textrm{mixed}}(p,r)
\ln \Big[\frac{(A+1)^{2}-(ab)^{2}}{(A-1)^{2}-(ab)^{2}}\Big]
+(p\leftrightarrow -r)
+(p\leftrightarrow r-p)\bigg]
\bigg\}, \hspace{.8cm}
\eeq
\beq
\label{vert34_ga}
\Ga^{(3)}_{\ga^{2}}(p,r)
& = & \frac{i\ga^{2}}{(4\pi)^{2}\theta^{4}}\,
\bigg\{
12\bigg[\frac{1}{\ep}+\ln\Big(\frac{\mu^{2}}{m^{2}}\Big)\bigg]
+\al^{(3)}_{\ga^{2}}(p,r)\ln(1+4b)
+18 
\nn
\\
&&
+\, \bigg[\be^{(3)}_{\ga^{2},\,\textrm{mixed}}(p,r)
\ln \Big[\frac{(A+1)^{2}-(ab)^{2}}{(A-1)^{2}-(ab)^{2}}\Big]
+\be^{(3)}_{\ga^{2},\,\textrm{light}}(p,r)
\ln\Big(\frac{1+c}{1-c}\Big)
\nn
\\
&&
+\, \be^{(3)}_{\ga^{2},\,\textrm{ghost}}(p,r)
\ln \Big(\frac{1+d}{1-d}\Big)
+(p\leftrightarrow -r)
+(p\leftrightarrow r-p)\bigg]
\bigg\}, \hspace{.8cm}
\eeq
where $\al$'s, $\be$'s and $\xi$'s are coefficients with polynomial
dependencies on the external momenta. The full explicit form of these
expressions are very bulky and we do not present them here. On the
other hand, since they are polynomials the corresponding contributions
are local and these explicit expression is not really important for our
analysis. Remember that the notations $a$, $b$, $A$, $c$, $d$ are
defined in \eqref{epsil}, \eqref{AA} and \eqref{cd}, respectively.

\begin{figure}[H]
\centering
\includegraphics[scale=0.19]{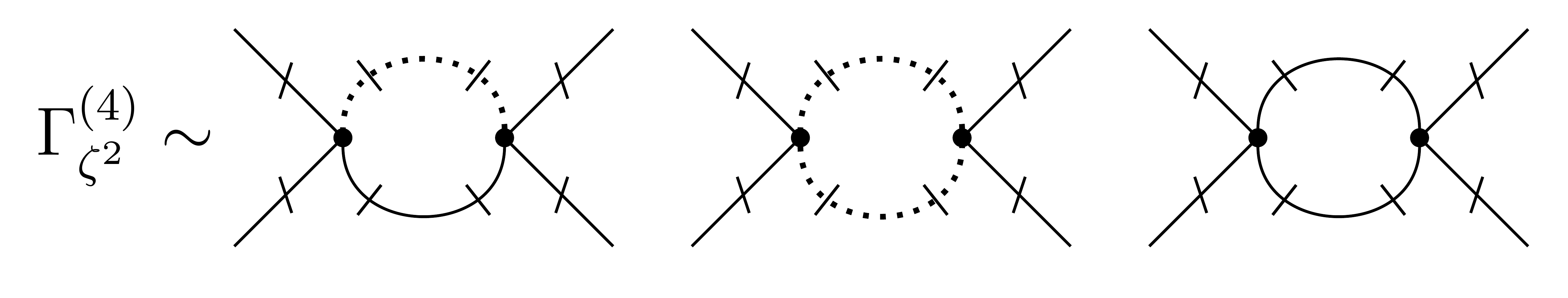}
\vspace{-3mm}
\caption{Diagrams for the four-point function with
the interaction vertex $\ze^{2}$.}
\label{fig12}
\end{figure}

\begin{figure}[H]
\centering
\includegraphics[scale=0.33]{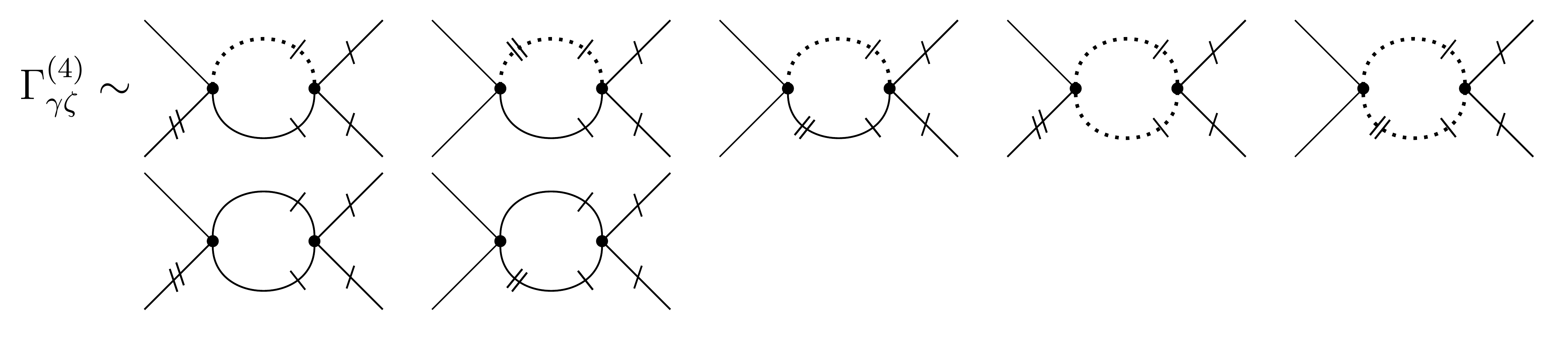}
\vspace{-3mm}
\caption{Diagrams for the four-point function with
the interaction vertex $\ga\ze$.}
\label{fig13}
\end{figure}

\begin{figure}[H]
\centering
\includegraphics[scale=0.33]{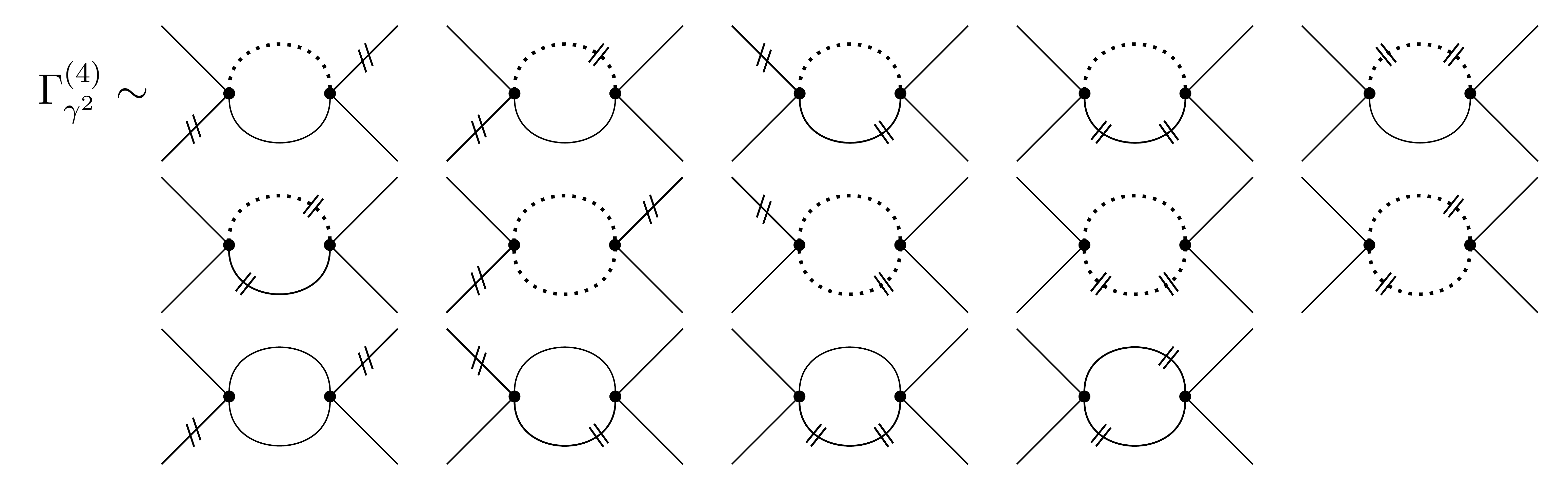}
\vspace{-3mm}
\caption{Diagrams for the four-point function with
the interaction vertex $\ga^{2}$.}
\label{fig14}
\end{figure}

It is important that the most essential, non-local parts of the
expressions have standard logarithmic structures, similar to those
already evaluated in section \ref{sec4}. Therefore, it should be
expected that the corrections above represent asymptotic
behavior in the IR, similar to the case of the propagator
corrections considered in the main part of the paper. Let us note
that we verified and confirmed the quadratic decoupling of the
heavy mode in the vertex terms. Furthermore, the same behavior
is observed for the four-point vertex corrections. In this case, the
diagrams of interest are depicted in Figures \ref{fig12}, \ref{fig13}
and \ref{fig14}.
The analytic expressions corresponding to these diagrams are
\beq
\label{Vert44_ze2}
\Ga^{(4)}_{\ze^{2}}(p,r,q)\big|^{(2\om)}
& = &
-\,\frac{8\ze^{2}}{\th^{4}(m^{2}-M^{2})^{2}}
\int \frac{d^{2\om}k}{(2\pi)^{2\om}}\,\Ga^{(4,4)}_{\ze^{2}}
\biggl\{\frac{2}{(k^{2}-m^{2})\big[(k-p)^{2}-M^{2}\big]}
\nn
\\
&& -\,\,
\frac{1}{(k^{2}-m^{2})\big[(k-p)^{2}-m^{2}\big]}
\,-\,\frac{1}{(k^{2}-M^{2})\big[(k-p)^{2}-M^{2}\big]}
\biggr\}
\nn
\\
&& +\,\, t\textrm{- and }u\textrm{-channel contributions},
\eeq
\beq
\label{Vert44_gaze}
\Ga^{(4)}_{\ga\ze}(p,r,q)\big|^{(2\om)}
& = &
\frac{4\ga\ze}{\th^{4}(m^{2}-M^{2})^{2}}\int \frac{d^{2\om}k}{(2\pi)^{2\om}}\,\Ga^{(4,4)}_{\ga\ze}
\biggl\{\frac{2}{(k^{2}-m^{2})\big[(k-p)^{2}-M^{2}\big]}
\nn
\\
&& -\,\,\frac{1}{(k^{2}-m^{2})\big[(k-p)^{2}-m^{2}\big]}
\,-\,\frac{1}{(k^{2}-M^{2})\big[(k-p)^{2}-M^{2}\big]}\biggr\}
\nn
\\
&& +\,\, t\textrm{- and }u\textrm{-channel contributions}, \eeq
\beq
\label{Vert44_ga2}
\Ga^{(4)}_{\ga^{2}}(p,r,q)\big|^{(2\om)}
& = &
-\,\frac{2\ga^{2}}{\th^{4}(m^{2}-M^{2})^{2}}\int \frac{d^{2\om}k}{(2\pi)^{2\om}}\,\Ga^{(4,4)}_{\ga^{2}}
\biggl\{\frac{2}{(k^{2}-m^{2})\big[(k-p)^{2}-M^{2}\big]}
\nn
\\
&& -\,\,\frac{1}{(k^{2}-m^{2})\big[(k-p)^{2}-m^{2}\big]}
\,-\,\frac{1}{(k^{2}-M^{2})\big[(k-p)^{2}-M^{2}\big]}\biggr\}
\nn
\\
&& +\,\, t\textrm{- and }u\textrm{-channel contributions},
\eeq
where, for the $s$-channel diagrams,
\beq
\label{vf44}
&&\Ga^{(4,4)}_{\ze^{2}} \,=\,
\big[r^{2}(k\cdot q)
-k^2(r\cdot q)
+2(k\cdot q) \big(r\cdot q\big)
+q^{2}(k\cdot r)
-2(k\cdot r)\big(k\cdot q-r\cdot q\big)\big]
\nn
\\
&&\hspace{1.45cm}
\,\,\times \big[k^2 \big(p^2-p\cdot r-p\cdot q\big)
-p^{2}(k\cdot r+k\cdot q)
+2(k\cdot r+k\cdot q)(p\cdot r+p\cdot q)
\nn
\\
&&\hspace{1.45cm}
\,\,+(r^2+q^2)(k\cdot p)
-2(k\cdot p)\big(k\cdot r+k\cdot q
+p\cdot r+p\cdot q-r\cdot q\big)
+2 (k\cdot p)^2\big],
\nn
\\
&&\Ga^{(4,4)}_{\ga\ze} \,=\,
-2 \big[k^2(r\cdot q)
-r^{2}(k\cdot q)
-2(k\cdot q) \big(r\cdot q\big)
+2(k\cdot r)\big(k\cdot q-r\cdot q\big)
-q^{2}(k\cdot r)\big]
\nn
\\
&&\hspace{1.45cm}
\,\,\times \big[k^2-k\cdot r-k\cdot q
+p^2-p\cdot r-p\cdot q+r^2
+2(r\cdot q)+q^2\big],
\nn
\\
&&\Ga^{(4,4)}_{\ga^{2}} \,=\,
4\big[k^2-k\cdot r-k\cdot q+p^2
-p\cdot r-p\cdot q+r^2
+2(r\cdot q)+q^2\big]
\nn
\\
&&\hspace{1.45cm}
\,\,\times \big(k^2-k\cdot r-k\cdot q+r^2+r\cdot q+q^2\big).
\eeq
The analytic expressions of the four-point vertex corrections
involving the couplings $\ze^{2}$, $\ga\ze$ and $ \ga^{2} $ are,
respectively,
\beq
\label{vert44_ze}
\Ga^{(4)}_{\ze^{2}}(p,r,q)
& = & -\frac{i\ze^{2}}{(4\pi)^{2}\theta^{4}}\,
\bigg\{
20\big[p^{2}(r\cdot q)-r^{2}(p\cdot q)
-2(p\cdot q)(r\cdot q)-q^{2}(p\cdot r)
\nn
\\
&&
+\, 2(p\cdot r)\big(p\cdot q-r\cdot q\big)\big]
\bigg[\frac{1}{\ep}
+\ln\Big(\frac{\mu^{2}}{m^{2}}\Big)\bigg]
+\al^{(4)}_{\ze^{2}}(p,r,q)\ln(1+4b)
\nn
\\
&&
+\, \be^{(4,\,s)}_{\ze^{2},\,\textrm{mixed}}(p,r,q)
\ln \Big[\frac{(A+1)^{2}-(ab)^{2}}{(A-1)^{2}-(ab)^{2}}\Big]
\bigg|_{p\leftrightarrow r+q}
+\xi^{(4)}_{\ze^{2}}(p,r,q)
\nn
\\
&&
+\,\be^{(4,\,s)}_{\ze^{2},\,\textrm{light}}(p,r,q)
\ln\Big(\frac{1+c}{1-c}\Big)\bigg|_{p\leftrightarrow r+q}
+\be^{(4,\,s)}_{\ze^{2},\,\textrm{ghost}}(p,r,q)
\ln \Big(\frac{1+d}{1-d}\Big)\bigg|_{p\leftrightarrow r+q}
\nn
\\
&&
+\, \be^{(4,\,t)}_{\ze^{2},\,\textrm{light}}(p,r,q)
\ln\Big(\frac{1+c}{1-c}\Big)\bigg|_{p\leftrightarrow q-p}
+\be^{(4,\,t)}_{\ze^{2},\,\textrm{ghost}}(p,r,q)
\ln \Big(\frac{1+d}{1-d}\Big)\bigg|_{p\leftrightarrow q-p}
\nn
\\
&&
+\, \be^{(4,\,t)}_{\ze^{2},\,\textrm{mixed}}(p,r,q)
\ln \Big[\frac{(A+1)^{2}-(ab)^{2}}{(A-1)^{2}-(ab)^{2}}\Big]
\bigg|_{p\leftrightarrow q-p}
\nn
\\
&&
+\, \be^{(4,\,u)}_{\ze^{2},\,\textrm{light}}(p,r,q)
\ln\Big(\frac{1+c}{1-c}\Big)\bigg|_{p\leftrightarrow r-p}
+\be^{(4,\,u)}_{\ze^{2},\,\textrm{ghost}}(p,r,q)
\ln \Big(\frac{1+d}{1-d}\Big)\bigg|_{p\leftrightarrow r-p}
\nn
\\
&&
+\, \be^{(4,\,u)}_{\ze^{2},\,\textrm{mixed}}(p,r,q)
\ln \Big[\frac{(A+1)^{2}-(ab)^{2}}{(A-1)^{2}-(ab)^{2}}\Big]
\bigg|_{p\leftrightarrow r-p}
\bigg\},
\eeq
\beq
\label{vert44_gaze}
\Ga^{(4)}_{\ga\ze}(p,r,q)
& = & \frac{i\ga\ze}{(4\pi)^{2}\theta^{4}}\,
\bigg\{
12(p\cdot r+p\cdot q-r\cdot q)\bigg[\frac{1}{\ep}+\ln\Big(\frac{\mu^{2}}{m^{2}}\Big)\bigg]
+\xi^{(4)}_{\ga\ze}(p,r,q)
\nn
\\
&&
+\, \al^{(4)}_{\ga\ze}(p,r,q)\ln(1+4b)
+\be^{(4,\,s)}_{\ga\ze,\,\textrm{light}}(p,r,q)
\ln\Big(\frac{1+c}{1-c}\Big)\bigg|_{p\leftrightarrow r+q}
\nn
\\
&&
+\, \be^{(4,\,s)}_{\ga\ze,\,\textrm{mixed}}(p,r,q)
\ln \Big[\frac{(A+1)^{2}-(ab)^{2}}{(A-1)^{2}-(ab)^{2}}\Big]
\bigg|_{p\leftrightarrow r+q}
\nn
\\
&&
+\, \be^{(4,\,s)}_{\ga\ze,\,\textrm{ghost}}(p,r,q)
\ln \Big(\frac{1+d}{1-d}\Big)\bigg|_{p\leftrightarrow r+q}
+\be^{(4,\,t)}_{\ga\ze,\,\textrm{light}}(p,r,q)
\ln\Big(\frac{1+c}{1-c}\Big)\bigg|_{p\leftrightarrow q-p}
\nn
\\
&&
+\, \be^{(4,\,t)}_{\ga\ze,\,\textrm{mixed}}(p,r,q)
\ln \Big[\frac{(A+1)^{2}-(ab)^{2}}{(A-1)^{2}-(ab)^{2}}\Big]
\bigg|_{p\leftrightarrow q-p}
\nn
\\
&&
+\, \be^{(4,\,t)}_{\ga\ze,\,\textrm{ghost}}(p,r,q)
\ln \Big(\frac{1+d}{1-d}\Big)\bigg|_{p\leftrightarrow s-p}
+\be^{(4,\,u)}_{\ga\ze,\,\textrm{light}}(p,r,q)
\ln\Big(\frac{1+c}{1-c}\Big)\bigg|_{p\leftrightarrow r-p}
\nn
\\
&&
+\, \be^{(4,\,u)}_{\ga\ze,\,\textrm{mixed}}(p,r,q)
\ln \Big[\frac{(A+1)^{2}-(ab)^{2}}{(A-1)^{2}-(ab)^{2}}\Big]
\bigg|_{p\leftrightarrow r-p}
\nn
\\
&&
+\, \be^{(4,\,u)}_{\ga\ze,\,\textrm{ghost}}(p,r,q)
\ln \Big(\frac{1+d}{1-d}\Big)\bigg|_{p\leftrightarrow r-p}
\bigg\}
\eeq
and
\beq
\label{vert44_ga}
\Ga^{(4)}_{\ga^{2}}(p,r,q)
& = & \frac{i\ga^{2}}{(4\pi)^{2}\theta^{4}}\,
\bigg\{
24\bigg[\frac{1}{\ep}+\ln\Big(\frac{\mu^{2}}{m^{2}}\Big)\bigg]
+\al^{(4)}_{\ga^{2}}(p,r,q)\ln(1+4b)
+36 
\nn
\\
&&
+\, \be^{(4,\,s)}_{\ga^{2},\,\textrm{light}}(p,r,q)
\ln\Big(\frac{1+c}{1-c}\Big)\bigg|_{p\leftrightarrow r+q}
+\be^{(4,\,s)}_{\ga^{2},\,\textrm{ghost}}(p,r,q)
\ln \Big(\frac{1+d}{1-d}\Big)\bigg|_{p\leftrightarrow r+q}
\nn
\\
&&
+\, \be^{(4,\,s)}_{\ga^{2},\,\textrm{mixed}}(p,r,q)
\ln \Big[\frac{(A+1)^{2}-(ab)^{2}}{(A-1)^{2}-(ab)^{2}}\Big]
\bigg|_{p\leftrightarrow r+q}
\nn
\\
&&
+\, \be^{(4,\,t)}_{\ga^{2},\,\textrm{light}}(p,r,q)
\ln\Big(\frac{1+c}{1-c}\Big)\bigg|_{p\leftrightarrow q-p}
+\be^{(4,\,t)}_{\ga^{2},\,\textrm{ghost}}(p,r,q)
\ln \Big(\frac{1+d}{1-d}\Big)\bigg|_{p\leftrightarrow q-p}
\nn
\\
&&
+\, \be^{(4,\,t)}_{\ga^{2},\,\textrm{mixed}}(p,r,q)
\ln \Big[\frac{(A+1)^{2}-(ab)^{2}}{(A-1)^{2}-(ab)^{2}}\Big]
\bigg|_{p\leftrightarrow q-p}
\nn
\\
&&
+\, \be^{(4,\,u)}_{\ga^{2},\,\textrm{light}}(p,r,q)
\ln\Big(\frac{1+c}{1-c}\Big)\bigg|_{p\leftrightarrow r-p}
+\be^{(4,\,u)}_{\ga^{2},\,\textrm{ghost}}(p,r,q)
\ln \Big(\frac{1+d}{1-d}\Big)\bigg|_{p\leftrightarrow r-p}
\nn
\\
&&
+\, \be^{(4,\,u)}_{\ga^{2},\,\textrm{mixed}}(p,r,q)
\ln \Big[\frac{(A+1)^{2}-(ab)^{2}}{(A-1)^{2}-(ab)^{2}}\Big]
\bigg|_{p\leftrightarrow r-p} \bigg\}.
\eeq
The indices in the coefficients $\be^{(4)}(p,r,q)$ denote $s-,\,
t-$ and $u-$channel contributions.

It is easy to note that these expressions are in a good qualitative
agreement with the self-energy corrections and ones  for the
three-point vertices.



\begin{thebibliography}{99}

\bibitem{antmot}  I. Antoniadis and E. Mottola,
{\it Four-dimensional quantum gravity in the conformal sector,}
Phys. Rev. {\bf D45} (1992) 2013.

\bibitem{apco} E.V. Gorbar and I.L. Shapiro,
\textit{Renormalization group and decoupling in curved space,}
JHEP {\bf 02} (2003) 021,
hep-ph/0210388.

\bibitem{DCCrun} I.L.~Shapiro, J.~Sol\`{a},
{\it On the possible running of the cosmological 'constant'},
Phys. Lett. {\bf B682} (2009)  105,   hep-th/0910.4925.

\bibitem{AC} T. Appelquist and J. Carazzone,
{\it Infrared singularities and massive fields,}
Phys. Rev. {\bf D11} (1975) 2856.

\bibitem{fervi} E.V. Gorbar and I.L. Shapiro,
\textit{Renormalization group and decoupling in curved
space, II. The standard model and beyond,}
JHEP {\bf 06} (2003) 004,
hep-ph/0303124.

\bibitem{Codello} A.~Codello and O.~Zanusso,
{\it On the non-local heat kernel expansion,}
J. Math. Phys.  {\bf 54} (2013) 013513,
arXiv:1203.2034.

\bibitem{Omar-FF4D}
S.A.~Franchino-Viñas, T.~de Paula Netto, I.L.~Shapiro,
and O.~Zanusso,
{\it Form factors and decoupling of matter fields in four-dimensional
gravity,}
Phys. Lett. {\bf B790} (2019) 229,
arXiv:1812.00460;
\\
S. A. Franchino-Viñas, T. de Paula Netto, and O. Zanusso,
{\it Vacuum effective actions and mass-dependent renormalization
in curved space,}
Universe {\bf 5} (2019) 
67,
arXiv:1902.03167.

\bibitem{Sebastian} S.A. Franchino-Viñas,
{\it Resummed heat-kernel for surface contributions: Dirichlet
semitransparent boundary conditions,}
arXiv:2208.11979. 

\bibitem{bexi} G.B. Peixoto, E.V. Gorbar and I.L. Shapiro,
{\it On the renormalization group for the interacting
massive scalar field theory in curved space,}
Class. Quant. Grav. {\bf 21} (2004) 2281, 
hep-th/0311229.

\bibitem{OdSh-91} S.D.~Odintsov and I.L.~Shapiro,
\textit{Perturbative approach to induced quantum gravity,}
Class. Quant. Grav. \textbf{8} (1991), L57-60.

\bibitem{rie} R.J.~Riegert,
{ \it A non-local action for the trace anomaly},
Phys. Lett. {\bf B134} (1984) 56. 

\bibitem{frts84} E.S. Fradkin and A.A. Tseytlin,
{\it Conformal anomaly in Weyl theory and anomaly free
superconformal theories,}
Phys. Lett. {\bf B134} (1984) 187.

\bibitem{duff94} M.J. Duff,
{\it Twenty years of the Weyl anomaly,}
Class. Quant. Grav. {\bf 11} (1994) 1387,
hep-th/9308075.

\bibitem{OUP} I.L. Buchbinder and I.L. Shapiro,
\textit{Introduction to Quantum Field Theory with Applications
to Quantum Gravity} (Oxford University Press, 2021).

\bibitem{Stelle77} K.S. Stelle,
{ \it Renormalization of higher derivative quantum gravity},
Phys. Rev. {\bf D16} (1977) 953.

\bibitem{highderi} M. Asorey, J.L. L\'opez, and I.L. Shapiro,
{\it Some remarks on high derivative quantum gravity,}
Int. Journ. Mod. Phys. {\bf A12} (1997) 5711,
hep-th/9610006. 

\bibitem{frts82} E.S. Fradkin and  A.A. Tseytlin,
{\it Renormalizable asymptotically free quantum theory of gravity,}
Nucl. Phys. {\bf B201} (1982) 469.

\bibitem{avbar86} I.G. Avramidi and A.O. Barvinsky,
{\it Asymptotic freedom In higher derivative quantum gravity,}
Phys. Lett. {\bf B159} (1985) 269.  

\bibitem{a} I.L. Shapiro and A.G. Jacksenaev,
{ \it Gauge dependence in higher derivative quantum
gravity and the conformal anomaly problem},
Phys. Lett. {\bf B324} (1994) 286.

\bibitem{SRQG-betas} L.~Modesto, L.~Rachwa\l, \ 
I.L.~Shapiro,
{\it Renormalization group in super-renormalizable quantum gravity,}
Eur. Phys. J. {\bf C78} (2018) 555,
arXiv:1704.03988.

\bibitem{Don-94} J.F. Donoghue,
\textit{Leading quantum correction to the Newtonian potential,}
Phys. Rev. Lett. {\bf 72} (1994) 2996,
gr-qc/9310024;
\textit{General relativity as an effective field theory: The leading
quantum corrections,}
Phys. Rev. {\bf D50} (1994) 3874,
gr-qc/9405057.

\bibitem{Gauss} G. de Berredo-Peixoto and I.L. Shapiro,
{\it Higher derivative quantum gravity with Gauss-Bonnet term,}
Phys. Rev. {\bf D71} (2005)  064005,
hep-th/0412249.

\bibitem{Polemic} I.L. Shapiro,
\textit{Polemic notes on IR perturbative quantum gravity,}
Int. J. Mod. Phys. {\bf A24} (2009) 1557,
arXiv:0812.3521.

\bibitem{Mottola-2017} E.~Mottola,
\textit{Scalar gravitational waves in the effective theory
of gravity,}
JHEP \textbf{07} (2017) 043;
Erratum: JHEP \textbf{09} (2017) 107,
arXiv:1606.09220. 

\bibitem{anomaly-2004} M. Asorey, E.V. Gorbar and I.L. Shapiro,
\textit{Universality and ambiguities of the conformal anomaly,}
Class. Quant. Grav. {\bf 21} (2004) 163,
hep-th/0307187.

\bibitem{BoxAno} M. Asorey, G. de Berredo-Peixoto and I.L. Shapiro,
\textit{Renormalization ambiguities and conformal anomaly in
metric-scalar backgrounds,}
Phys. Rev. {\bf D74} (2006) 124011,
hep-th/0609138.

\bibitem{AnInt22}
M. Asorey, W. Cesar e Silva, I.L. Shapiro and P.R.B. do Vale,
{\it Trace anomaly and induced action for a metric-scalar background,}
arXiv:2202.00154.

\bibitem{Poly81} A.M. Polyakov,
{\it Quantum geometry of bosonic strings,}
Phys. Lett. {\bf B103} (1981) 207.

\bibitem{induce}
I.L.~Shapiro and G.~Cognola,
\textit{Interaction of low-energy induced gravity with quantized
matter and phase transition induced to curvature,}
Phys. Rev. \textbf{D51} (1995) 2775, 
hep-th/9406027.
\bibitem{AntMazMot97} I. Antoniadis, P.O. Mazur and E. Mottola,
\textit{Physical states of the quantum conformal factor,}
Phys. Rev. \textbf{D55} (1997)  4770, hep-th/9509169.

\bibitem{bavi85}  A.O. Barvinsky and G.A. Vilkovisky,
{\it The generalized Schwinger-DeWitt technique in gauge theories
and quantum gravity,}
Phys. Repts. {\bf 119} (1985) 1.

\bibitem{AntoOdin94} I. Antoniadis and S.D. Odintsov,
{\it Renormalization group and logarithmic corrections to scaling
relations in the conformal sector of 4D gravity,}
Phys. Lett. {\bf B343} (1995) 76,
arXiv:hep-th/9411012.

\bibitem{Holdom} B. Holdom,
{\it Running couplings and unitarity in a 4-derivative scalar field theory,}
Phys. Lett.  \textbf{B843} (2023) 138023,
arXiv:2303.06723.

\bibitem{Cremin} P. Creminelli, A. Nicolis, M. Papucci and
E. Trincherini,
{\it Ghosts in massive gravity,}
JHEP {\bf 0509} (2005) 003,
hep-th/0505147.

\bibitem{julton} J. Julve and M. Tonin,
{\it Quantum gravity with higher derivative terms,}
Nuovo Cim. {\bf B46} (1978) 137.  

\bibitem{Leibb75} G. Leibbrandt, {\it Introduction to the
Technique of Dimensional Regularization,}
Rev. Mod. Phys. \textbf{47} (1975) 849.

\bibitem{Patel} H.H. Patel,
{\it Package-X: A Mathematica package for the analytic
calculation of one-loop integrals,}
Comput. Phys. Commun. {\bf 197} (2015) 276,
arXiv:1503.01469.

\bibitem{Wolfram} Wolfram Research, Inc.,
\textit{Mathematica,} (Version 12.0, Champaign, IL, 2019).

\bibitem{Collins} J.C. Collins,
\textit{Renormalization} (Cambridge University Press, 1984).

\bibitem{PeskSchr} M.E. Peskin and D.V. Schroeder,
\textit{An Introduction to Quantum Field Theory} (Westview Press, 1995).

\bibitem{Burgess}  C.P.~Burgess,
{\it Quantum gravity in everyday life: General relativity as an
effective field theory,}
Living Rev. Rel. {\bf 7} (2004) 5,
gr-qc/0311082.

\bibitem{2simpQG} E.V. Gorbar and I.L. Shapiro,
\textit{Nonlocality of quantum matter corrections and cosmological
constant running,}
JHEP \textbf{2022} (2022) 103, 
arXiv:2203.09232.

\bibitem{don22} J.F. Donoghue,
\textit{Nonlocal partner to the cosmological constant,}
Phys. Rev. \textbf{D105} (2022) 10, 105025
arXiv: 2201.12217.

\bibitem{Weinberg89}  S. Weinberg,
\textit{The cosmological constant problem,}
Rev. Mod. Phys. {\bf 61} (1989) 1.

\bibitem{CC-nova} I.L. Shapiro, J. Sol\`{a},
{\it Scaling behavior of the cosmological constant:
Interface between quantum field theory and cosmology,}
JHEP {\bf 02} (2002) 006,
hep-th/0012227.

\bibitem{Padm-03} T. Padmanabhan,
\textit{Cosmological constant—the weight of the vacuum,}
Phys. Rep. {\bf 380} (2003) 235,
arXiv:hep-th/0212290.

\bibitem{SahniStar-2006} V.~Sahni and A.A.~Starobinsky,
\textit{Reconstructing Dark Energy,}
Int. J. Mod. Phys. D \textbf{15} (2006) 2105, 
astro-ph/0610026. 

\bibitem{AntoMzMtt94} I. Antoniadis, P.O. Mazur and E. Mottola,
{\it Scaling behavior of quantum four-geometries,}
Phys. Lett. {\bf B323} (1994) 284,
arXiv:hep-th/9301002.

\bibitem{Eliz94_OneLoop}
E. Elizalde, A.G. Jacksenaev, S.D. Odintsov and I.L. Shapiro,
{\it One-loop renormalization and asymptotic behaviour of a
higher-derivative scalar theory in curved spacetime,}
Phys. Letters {\bf B328} (1994) 297,
hep-th/9402154;
{\it A four-dimensional theory for quantum gravity with
conformal and non-conformal explicit solutions,}
Class. Quant. Grav. {\bf 12} (1995) 1385,
hep-th/9412061.

\bibitem{SusyConf96} I.L. Buchbinder and A.Yu Petrov,
\textit{Quantum dynamics of N=1, D=4 supergravity chiral compensator,}
Class. Quant. Grav. {\bf 13} (1996) 2081,
arXiv:hep-th/9511205;
\textit{On quantum model of supergravity compensator,}
Mod. Phys. Lett. {\bf A11} (1996) 2159,
arXiv:hep-th/9604154.

\bibitem{FF-19} T.G. Ribeiro and I.L. Shapiro
\textit{ Scalar model of effective field theory in curved space,}
JHEP {\bf 2019} (2019) 163,
arXiv:1908.01937.

\bibitem{Sola22} J. Sol\`a Peracaula,
\textit{The cosmological constant problem and running
vacuum in the expanding universe,}
Phil. Trans. R. Soc. A. \textbf{380} (2022) 20210182,
arXiv:2203.13757.

\bibitem{Mottola22} E. Mottola,
\textit{The effective theory of gravity and dynamical vacuum energy,}
JHEP \textbf{2022} (2022) 37,
arXiv:2205.04703.

\bibitem{ABSh2} A.~Accioly, B.L.~Giacchini and I.L.~Shapiro,
{\it On the gravitational seesaw in higher-derivative gravity},
Eur. Phys. J.  {\bf C77} (2017) 540,
gr-qc/1604.07348.

\bibitem{Krasni_87} N.V. Krasnikov,
{\it Nonlocal gauge theories,}
Theor. Math. Phys. {\bf 73} (1987) 1184.

\bibitem{Kuz_89} Y.V. Kuz’min,
{\it The convergent nonlocal gravitation} (in Russian),
Yad. Fiz. {\bf 50} (1989) 1630;
[Sov. J. Nucl. Phys. {\bf 50} (1989) 1011].

\bibitem{Tombou_97} E.T. Tomboulis,
{\it Superrenormalizable gauge and gravitational theories,}
hep-th/9702146;
\
{\it Nonlocal and quasilocal field theories,}
Phys. Rev. {\bf D92} (2015) 125037,
arXiv:1507.00981.

\bibitem{Modesto-nonloc} L. Modesto,
{\it Super-renormalizable quantum gravity,}
Phys. Rev. {\bf D86} (2012) 044005,
arXiv:1107.2403;
\
L.~Modesto and L.~Rachwa\l,
{\it Super-renormalizable and finite gravitational theories,}
Nucl. Phys. {\bf B889} (2014)  228,
arXiv:1407.8036; \
{\it Nonlocal quantum gravity: A review,}
Int. J. Mod. Phys. {\bf D26} (2017) 1730020.

\bibitem{CountGhosts} I.L. Shapiro,
{\it Counting ghosts in the ``ghost-free'' non-local gravity.}
Phys. Lett. {\bf B744} (2015) 67,
arXive:1502.00106.

\end{thebibliography}
\end{document}